# Water on The Moon, III. Volatiles & Activity

Arlin Crotts (Columbia University)

For centuries some scientists have argued that there is activity on the Moon (or water, as recounted in Parts I & II), while others have thought the Moon is simply a dead, inactive world. [1] The question comes in several forms: is there a detectable atmosphere? Does the surface of the Moon change? What causes interior seismic activity?

From a more modern viewpoint, we now know that as much carbon monoxide as water was excavated during the *LCROSS* impact, as detailed in Part I, and a comparable amount of other volatiles were found. At one time the Moon outgassed prodigious amounts of water and hydrogen in volcanic fire fountains, but released similar amounts of volatile sulfur (or $SO_2$), and presumably large amounts of carbon dioxide or monoxide, if theory is to be believed. So water on the Moon is associated with other gases.

Astronomers have agreed for centuries that there is no firm evidence for "weather" on the Moon visible from Earth, and little evidence of thick atmosphere. [2] How would one detect the Moon's atmosphere from Earth? An obvious means is atmospheric refraction. As you watch the Sun set, its image is displaced by Earth's atmospheric refraction at the horizon from the position it would have if there were no atmosphere, by roughly 0.6 degree (a bit more than the Sun's angular diameter). On the Moon, any atmosphere would cause an analogous effect for a star passing behind the Moon during an occultation (multiplied by two since the light travels both into and out of the lunar atmosphere). Furthermore, since lunar surface gravity is one-sixth Earth's, its atmosphere is six times more extended. However, the Moon radius is one-quarter Earth's, so the path through its atmosphere is four times smaller. By timing the disappearance of an occulted star, 19th century observers found the star never deviated more than 1.5 arcsecond (about 0.0004 degree) from its predicted (vacuum) position, so atmospheric pressure at the Moon's surface must be less than $(0.0004°/0.6°) \times \frac{1}{2} \times \frac{1}{6} \times 4$ or about 1/4000th of Earth's atmospheric pressure (assuming both atmospheres have the same composition and temperature). [3] Modern occultation tests using the refraction of radio waves from spacecraft find densities at least 10 billion times smaller (for charged particles, especially electrons). [4]

It is clear that the Moon does not maintain enough atmosphere to support constantly changing activity like weather, but it still has an atmosphere. But is the Moon active? Studying the atmosphere will give us clues, but we will see that this is in part a controversial issue because it involves infrequent events, which can be problematic for the scientific method. But we will find that the Moon is not dead yet.

Placing instruments on the Moon is an effective way to answer some of these questions, and has been invaluable. To determine the Moon's atmospheric composition, mass spectrometers can be amazingly sensitive. The Lunar Atmospheric Composition Experiment (LACE) on *Apollo 17* could sense a few hundred molecules per cubic centimeter, or about 100 parts per quadrillion of atmospheric density at sea level on Earth. [5] An amazing result from LACE was the "wave atmosphere" of gas on the Moon, with a cyclic density profile moving around the globe, sweeping past each site once a month. As discussed in Part I, the cold lunar surface at night is sticky, and few molecules can fly freely in ballistic bounces once they hit this cold ground. Instead, the nighttime atmosphere is much thinner than during the day. The CCGE (Cold Cathode Gauge Experiment, or CCIG: Cold Cathode Ion Gauge – essentially the CCGE in different packaging) was placed on the Moon with ALSEP on *Apollo 12*, *14* and *15*, and measured the total number of particles hitting the detector: a density of ten million per cubic centimeter during the day, two hundred thousand at night. [6] At night these molecules were *ad*sorbed (onto the soil's surface, not *ab*sorbed inside). [7] At sunrise, however, the lunar soil heats and stuck atmospheric molecules vibrate free to enter the Moon's atmosphere again, bouncing over the lunar surface. For most atomic and molecular species, this is when the atmosphere is densest, about ten times the daytime average.

Millions of molecules per cubic centimeter and giant waves of gas sweeping east to west around the Moon every month seems impressive, but do not suffice to produce observable effects far from the lunar surface. But sunrise has other consequences. The sun is an intense ultraviolet source irradiating the lunar surface. Many solar photons are sufficiently energetic to knock electrons out of lunar soil grains via the photoelectric effect. Losing these electrons produces positive charges in sunlit areas, and charged dust grains are repelled from the surrounding charged soil to actually rise up and fly to less charged regions in dark areas. (Also, solar wind electrons can impart a negative charge in dark regions.) These electric fields are strongest at the boundary between light and dark, the local terminator, primarily at sunrise or sunset (but landing elsewhere around rocks, craters and local topographic relief). Thin clouds of dust hug the terminator, and are seen when illuminated at low sun angles, looking towards the rising or setting sun (Figure 1). While this horizon phenomenon was seen throughout the 1966–1968 Surveyor program (by *Surveyor 1, 5, 6* and *7*), these results went unexplained [8] and outside science journals, while similar observations of the solar corona (the Sun's extended atmosphere) were published. [9] Similar horizon glows were observed in 1973 by the robotic rover *Lunokhod 2* but were largely overlooked outside the Soviet Union. [10] In 1972 David Criswell of the Lunar Science Institute developed the theory of photoelectric dust levitation, and later that year Gene Cernan, Commander of *Apollo 17*, was asked to draw from lunar orbit the solar corona, to recreate the Surveyor observations. His results were shocking: not only could one see the corona, but a huge plume of light stretching far beyond the Moon, even beyond the "zodiacal light" – dust in orbit around the Sun within the inner Solar System. This implied dust somehow lifted tens of kilometers above the Moon, scattering sunlight toward the spacecraft. This was published (simultaneous to the formal Surveyor horizon glow paper) [11] but garnered little attention at the time. This huge light plume was solely a visually reported phenomenon, not seen on all missions (yes on *Apollo 8, 10, 15* and *17*, but not *Apollo 16*; with no observations attempted on *Apollo 11-14*). Results indicate a distribution of small dust particles, about 0.1 micron = 0.0001 millimeter across, spread over large distances from the Moon, and highly variable (but not explained by artificial contamination). It remained a mystery.

In 1994 *Clementine* ended a generation's hiatus in lunar exploration, carrying a type of imaging optical detector, CCDs (charge-coupled devices) now familiar in today's digital cameras but unavailable during Apollo. It could easily detect faint surface brightness signals like the lunar glow seen by Cernan, and found a weaker signal, which caused confusion but temporarily reinvigorated attention to the Moon's dust "atmosphere." [12] When new missions circa 2009 promised progress (for instance *Clementine*'s star tracker CCD had no filters, which limited analysis), interest increased yet again, especially when NASA decided to devote an entire mission to lunar dust and gas: the *Lunar Atmosphere & Dust Environment Explorer* (*LADEE*) scheduled for 2013 (more later in this Chapter). Inspired by this a model was developed for how dust particles are lofted many kilometers off the Moon.

This model of lofted lunar dust starts with the same photoelectric charging of the dust by solar ultraviolet light, but follows the fate of charged dust motes after that. [13] Put simply, the mote will not move until electrostatic forces overcome the sum of gravity and mechanical "sticking" forces. After coming unstuck, however, the mote accelerates away from the charged lunar surface with constant force. The velocity it attains depends on a quantity called the shielding or *Debye length*: the plasma that fills all space in the Solar System is composed of separate positive and negative charges, and, over the positively charged lunar surface, negative charges cluster closer to the ground until the surface's positive charge appears cancelled or "shielded" from the point of view of a distant charge. This distance is of order 10 meters at the terminator, and escaping charged dust motes (the smallest ones, about 0.01 micron) can attain velocities up to 1 kilometer per second, almost half lunar escape velocity, sufficient to loft them 100 kilometers. Note further electric fields reside in the solar wind, and perhaps some dust particles are even swept away from the Moon altogether by the solar wind.

These little dust storms likely have long-term consequences in moving dust mass. Normally shadowed areas in brighter plains (or vice versa) will collect (or lose) dust, presumably. Criswell's 1972 analysis implies typical dust loss of 1 milligram per square centimeter per year, which seems small but amounts to one tonne per square centimeter per billion years, more than the entire mass in lunar regolith and many times faster than micrometeorite erosion. Most regolith particles are too large to be lofted, and only dust at the surface feels the electrostatic force, so the same dust may move repeatedly. Long ago Thomas Gold was derided for hypothesizing that electrostatic forces might re-sculpt the lunar surface by moving large amounts of dust. [14] Perhaps he was half correct.

The interaction of the plasma around the Moon with its surface can be complex indeed, and lead to surprising consequences. Plasma in the solar wind can flow along the surface then over a crater rim and into the crater, leaving a wake of low plasma density just below the lip of the crater. Surprisingly, the void in the plasma left under this wake will grow a negative charge, since the electrons flow over the lip and into the void more easily than the heavier ions. [15] Charge particles can appear around the Moon at dawn and dusk due to the flow of material moving along Earth's magnetic field (its bow shock). [16] Most generally, the Moon is always influenced by the solar wind, which tends to strike the noonday surface at an angle perpendicular to the wind's entrained electric field. The wind's electric field direction tends to point from the sunrise side of the Moon to the sunset side, meaning that electrons far above the Moon tend to fly off in the direction of sunset, and positive ions towards sunrise. Once an atom or molecule picks up a charge, it leaves the Moon almost immediately. [17] This is not so surprising, considering that whole dust grains are flying from the Moon!

In a sense, the Moon has two coextensive but separate atmospheres: its ionized plasma and its electrically neutral gas, which rarely interact except when an atom or molecule is ionized or de-ionized (hence changing its membership among the two). The ionized atmosphere is quickly swept away (*advected*) from the Moon, while most neutral atoms and molecules simply bounce ballistically around the surface (until they take too fast a bounce and achieve escape velocity, or become ionized), occasionally sticking and unsticking from the surface and engaging in the monthly wave. The Moon is our local example of a general phenomenon: the *solid-bounded exosphere*, an atmosphere so thin that atoms and molecules rarely interact with each other, but do interact with the solid surface and with surrounding electric and magnetic fields. This type of atmosphere typifies nearly all objects in the Solar System, excepting Earth, Mars, Venus, Saturn's moon Titan, the gas giant planets, and to some degree Neptune's moon Triton. Mercury, Pluto, the asteroids and the many remaining moons of the planets have solid-bounded exospheres. We have the Moon nearby to tell us how the others work.

The neutral atmosphere is steadier since it is less affected by the solar wind, so let us continue to discuss what we know about it. Details about neutral gas come from LACE of *Apollo 17*'s ALSEP, introduced above. It operated for ten months in 1973, revealed the wave atmosphere effect, and measured the composition of the neutral atmosphere in many molecular and atomic species, from 1 to 110 atomic mass units (1 a.m.u. being roughly a proton or neutron's mass). [18] There are several comparable components of the atmosphere. Atomic hydrogen is surprisingly absent; the dominant form of hydrogen is molecular $H_2$ at roughly 60,000 per cubic centimeter. [19] Similar numbers of neon atoms are found, also coming from the solar wind, mostly neon-20 (the isotope with 20 neutrons and protons), but 7% of neon-22. Comparable numbers of helium-4 atoms are seen, from the solar wind plus a lesser amount from radioactive decay inside the Moon. In terms of mass, a large component is argon-40 from radioactive decay of potassium-40, but also some argon-36, common in the primordial composition of Solar System objects (argon of solar origin is 86% argon-36 versus 0.3% on Earth, the rest being argon-40). The day/night variation is evident: 10,000 versus 200 per cubic centimeter for argon, 40,000 versus 2000 for helium. [20]

The strongest signal for most species occurs near sunrise (and background noise is non-negligible), so this is the best time to look for rare species. Not all species stick well on the night side (such as $N_2$) and solar wind species do not necessarily behave this way, so the Moon's atmospheric composition changes depending on phase during the month. Nonetheless, LACE could look for these, and in sunrise number density the gases detected, excluding hydrogen, are – $^{40}$Ar: 20000, $^{4}$He: 15000, $^{36}$Ar: 1600±400, $CO_2$: 1400±1000, $CH_4$: 1200±400, $O_2$: 1000±1000, $N_2$: 800±800, CO: 800±800, $NH_3$: 600±600, $H_2O$: 600±600, in units of number per cubic centimeter, and with errors after the ± symbol being one standard deviation (usually 67% of the probability range of values). [21] In summary, argon-40 and helium-4 are strongly detected at about the same count level, but argon has ten times the mass and dominates the mass density. Argon-36 is detected at levels one might expect from the solar wind (about 800 atoms per cubic centimeter), but at only 8% the level seen for argon-40. If both types of argon originated in the solar wind, one would expect to see about 80 times less argon-40. Argon-40 must originate from lunar radioactive decay. Methane ($CH_4$) is significantly detected, and might originate from solar wind reactions in the lunar soil. Further molecules (carbon dioxide, oxygen, nitrogen, carbon monoxide, ammonia and water) are only marginally detected.

Three instruments provide various (conflicting?) measures of lunar atmospheric gas: SIDE (mass spectra: ions only – see Part I), LACE (mass spectra: neutrals only), and CCGE/CCIG (total particle count). [22] Later we see how this situation might improve. Suffice now to say that roughly equal numbers exist of hydrogen molecules, helium-4 and neon-20 atoms, and somewhat fewer argon-40, which is still significant in terms of mass. Summing all species, day or night, yields a total lunar atmospheric mass of about 20 tonnes (the mass of air in a 5000 square meter office building, the size of a soccer/football field, one storey high).

Argon-40 density varies wildly over a month, but also varies strongly from month to month. This density varied by a factor of two between consecutive months, and since a typical argon-40 atom is stored in the lunar atmosphere for about this long, flow from the argon-40 source must vary roughly 100% from month to month. Argon leaks from the Moon highly episodically. [23] Helium is also produced by radioactive decay, but even more is brought to the Moon by the solar wind. Argon-40 is a more useful tracer of the gas's origin, and the implications are surprising. On average total of several tonnes of argon reside in the lunar atmosphere, so about $2\times10^{21}$ atoms per second must be produced. This is a huge amount considering that potassium-40 is only about one hundred millionth part of the Moon's mass and decays with a long (1.25 billion year) half-life. This implies that a large fraction of the argon-40 produced by the Moon, about 8%, is leaking out, so some comes from deep inside. [24]

There are more ways to detect gas from radioactive decay, and this also indicates gas leaking episodically from the Moon. Most of these gas atoms come from isotopes heavier than potassium-40, particularly uranium-238 and thorium-232. [25] These decay by emitting helium-4 nuclei (alpha particles), which mix inconveniently with solar-wind helium in a confusing way. In a few steps uranium-238 decays to the gas radon-222, which is also radioactive and decays eventually to polonium-210, then finally to lead-206, a stable isotope. Radon-222 has a half-life of 3.8 days, long enough to be useful, whereas thorium-232 produces radon-220, with a 56-second half-life, decaying before it can do much.

In 3.8 days, can radon-222 leak to the lunar surface? Apparently it does. (We return to this.) Several times in orbit over the lunar surface, spacecraft have looked down to see a pool of radon-222 gas on the ground as they fly over. Radon is so heavy that it barely disperses, perhaps one hundred kilometers before it decays. (A radon atom's ballistic bounce is only a few hundred meters.) At least two such events occurred during *Apollo 15* and at least two more for *Lunar Prospector*. These and the *Kaguya* Alpha-Ray Detector all detected emission from the Aristarchus region. [26] These were detected with alpha particle spectrometers, which measure the energy of those helium-4 nuclei emitted by decaying nuclei. These alpha particles do not have the energy to penetrate even a piece of paper, but they can fly up from the Moon through the vacuum of space to be intercepted by a detector in orbit. By measuring their energy, we know the type of atomic nucleus from which they came. Half of the time when a radon-222 nucleus decays it will fire an alpha particle into space (and recoil into the ground were it will stay as polonium, bismuth and finally lead), and half the time the alpha will shoot into the ground and rocket the radon (now polonium) faster than lunar escape velocity, out into the Solar System. [27] It is the alphas from the first kind of decay geometry that the alpha particle spectrometers detect.

Since the radon-222 will disappear within a matter of weeks, these concentrations of gas must be fresh. Since the gas distribution differs when observed at various times, the events do not indicate continuous leakage. Still, the amount of radon-222 required to produce these signals is less than a gram per event, assuming all of the gas finds its way to the surface. Laboratory experiments in vacua using regolith simulants shows that radon does not move easily through the lunar soil, which implies that either much larger amounts of radon are being released into the regolith but not reaching the vacuum, or perhaps other gases are released and carry the radon through the regolith and into space. [28] Radon outgassing from the Moon appears episodic, and argon (a much larger contribution) enters the atmosphere in an off-and-on way. The Moon shows strong evidence of venting much of the gas around it in short-lived events.

Sufficient gas leaking through the soil can create a situation in which the whole layer of regolith is unstable to the gas explosively expanding into the vacuum. Lunar regolith is tremendously impermeable to gas, so even a small flow (less than 1 gram per second) will accumulate. Even a tonne of typical lunar gas (say, atomic mass 20 a.m.u., intermediate between 4 for helium and 40 for argon) creates a situation in which an explosion is energetically favored, blowing a plug of regolith into space as a dust cloud, which will expand to several kilometers in radius, taking a few minutes to do so. If gas interacts with the soil like this, by delivery at moderate or high flow rates to a small areas at the base of the regolith, several of these explosions might be expected to take place per month, considering that roughly ten tonnes per month leak out of the Moon. [29]

An important coincidence is that phenomena resembling this have been reported. While the term "transient lunar phenomena" (TLPs, or LTPs – see [55]) is a catch-all category for any temporary change in the appearance of the Moon (or even for its gas release events), the range of reported appearance is more limited. Over a thousand such reports, collected by the exhaustive (and exhausting) work of Winifred Sawtell Cameron and of Barbara M. Middlehurst, [30] and most are "brightenings" – white or color-neutral increases in surface brightness, "reddish" – red, orange or brown color changes with or without brightening, "bluish" – blue, blue-green or violet color changes with or without brightening, or "gaseous" – obscurations, misty or darkening changes in surface appearance. Nearly all TLPs are highly localized, usually to a radius much less than 100 km, often as unresolved points (corresponding to roughly 1 km or less). Aristarchus composes the site of a full quarter of all reports, and other sites on the Aristarchus Plateau (Cobra's Head, Schröter's Vally and Herodotus) add another 8%, placing almost one-third of all reports on the Plateau, an area of only 0.001% of the lunar surface. There are roughly another twenty sites each contributing 1% or more of the remaining reports, and another 80 or so sites with only 1 – 3 reports each. A map of the lunar near side with TLP report site frequency is shown in Figure 2 and Table 1.

At this point some readers have started to groan, because TLPs have a bad reputation in select audiences. By nature TLPs are anecdotal information, which is an issue of difficulty within the scientific method because such data are not subject to repeat experiment. The subjectivity of many TLP reports does not engender trust for their reliability. Establishing their consistent behavior is a major task required to assess their applicability in understand physics processes on the Moon. What is the consistent behavior of these reports? This requires a lengthy historical discussion and statistical analysis that some readers may wish to skip. This analysis is relocated to Appendix 1, which the reader should scrutinize. Several topics are consigned there: observations of meteoroid impacts on the Moon, naked-eye TLPs (including those seen by astronauts orbiting the Moon), TLP photographs, and emission of light by atoms in the lunar atmosphere. We will continue here with the results from this TLP analysis.

By "consistent" we mean behavior of TLPs independent of factors that characterize various properties of the observers, separate from lunar physics. By requiring TLP sites to maintain consistent behavior in the report catalog regardless of when or where the observations were made, the following features survive (Table A.9; with a statistical confidence of 99% or better): Aristarchus Plateau 46.7%±3.3%, Plato 15.6%±1.9%, Mare Crisium 4.1%±1.0%, Tycho 2.8%±0.8%, Kepler 2.1%±0.7%, Grimaldi 1.6%±0.6%, Copernicus 1.4%±0.6% (errors are ±1 standard deviation). Portraits of the features are shown in Figure 3, and we discuss each. Aristarchus and vicinity is responsible for roughly half of the reports and worthy of much more discussion. Four of the sites form the youngest, largest impacts on the near side: Aristarchus, Tycho, Kepler and Copernicus. Aristarchus seems in a class in itself; still the other three fresh craters together sum to 6.2%±1.2% of the reports. Plato and Grimaldi are both old, flooded craters at the edge of large mare areas (Imbrium and Procellarum, respectively). Mare Crisium is a unique, and troublesome, case.

The utility of this list is its likely greater reliability in comparison to other physical phenomena that may (or might not) be related; at least it is consistent and varies less with observer parameters. The physical phenomenon that is localizable and known to be related to lunar outgassing is radon alpha-particle events. The correspondence to the TLP list is marked: all four prominent radon-222 outgassing events land on this list's features: Aristarchus (twice), Grimaldi and Kepler. Indeed, this list of four events is entirely consistent with a random draw from the frequencies listed above for robust TLPs, especially considering that *Apollo 15* could not fly over Plato (or Tycho). The chances of these four events chosen at random and matching the robust TLP list are miniscule. Even a generous estimate for the area associated with each TLP site means that the probability of such a selection at random is about 0.0001; a more reasonable estimate is 30 times less than this. Almost certainly there is some reason these outgassing events are associated with robust TLP sites; the most plausible explanation is that there is some physical association between TLPs and outgassing.

Another way of measuring radon-222 outgassing activity is by counting alpha particles from its daughter product polonium-210, which is not produced in other significant ways. Even if radon-222 leaks out slowly, the accumulation of such activity over the past century or so will appear as a polonium-210 alpha particle decay signal at that location. A map of the polonium-210 alpha flux shows a notable correlation of the isotope with the boundary between lunar maria and highlands. [31] The polonium-210 distribution was also measured on *Lunar Prospector*, and this also had a high correlation with mare edges, with a probability of $6.5 \times 10^{-5}$ of being random. [32]

TLPs are obviously correlated with maria boundaries, as even a casual inspection of Figure 2 reveals. The robust sample (Figure 4) also correlates with maria edges with roughly 99.9% to 99.99% confidence level, including all events, robust or not, elevates this to a level that is essentially certain. [33] In addition to strong sources of TLPs correlating with sites of episodic radon-222 outgassing, even low-level TLP activity correlates with the same mare/highlands boundary locus that the long term radon outgassing traced by polonium-210 does.

TLPs correlation with outgassing in these two different ways lends support to the idea that they are physically tied to the Moon, not some observer-physiological or otherwise Earthbound effect. What is the nature of this association? Is it causative or merely correlational? What sort of outgassing effects might be expected, and could they plausibly cause TLPs. Remember that argon-40 appears to outgas episodically as well as radon-222, a tracer, so plausibly most of the mass in the lunar atmosphere has entered from the interior in discrete events.

Another physical association with the mare edges is moonquakes (which we discussed in Part I in connection with phase changes connected to water). While there is some correlation of shallow moonquakes with mare edges, the correspondence of deep (700 –1200 kilometer below surface) to mare edges is obvious (Figure 5). Few TLPs have actually been tied temporally, not convincingly more than random. Perhaps this is not surprising given the depth of quakes that correlate spatially with TLPs. TLPs correlate weakly with times of maximal tidal stress. Separately, moonquakes have been tied to episodic $^{40}$Ar release. [34]

On the Moon regolith is essentially ubiquitous, over nearly every surface to depths of perhaps 40 meters north of the South Pole-Aitken Basin on the Far Side to only 1 meter near Flamsteed (*Surveyor 1* site) and Lichtenberg craters in western and northern Oceanus Procellarum. How will outgassing interact with the regolith? This depends on the gas flow rate. At low rates, gas interacts with regolith on essentially a single molecular basis and percolates to the vacuum with little disturbance. At higher rates, regolith can "bubble" or fluidize; [35] yet higher, roughly $10^{24}$ molecules per second or more, gas will establish an overpressure relieved explosively by blowing through the regolith. At even higher flows one expects a jet of gas at approximately sonic velocities.

In some cases the outgassing event should leave a topographic feature that is resolvable at 1 meter resolution typical of *Lunar Orbiter III* and *V*, and the *Lunar Reconnaissance Orbiter* LROC-NAC images. Such an outburst essentially produces a crater of order 20 meters across and a spray of material extending over hundreds of meters (and more thinly over kilometers). [36] Whether we can detect these changes and the ones listed immediately above is in large part an issue distinguishing between regolith that has been residing on the surface, and that which has been protected from weathering effects of exposure to space.

Lunar regolith contains an appreciable amount of iron (typically 10–15%), and the surface state $Fe^{2+}$ of iron is responsible for strong absorption bands at 0.95 and 1.8 microns wavelength. This state is damaged gradually by exposure to micrometeorite impacts, which ultimately replace the surface states with a rim of glass containing $Fe^0$ rather than $Fe^{2+}$, and the absorption band disappears. Formally this is measured with OMAT (Optical MATurity parameter) based on the earliest comprehensive data set of lunar surface photometry, from *Clementine* in 1994, which included a band centered on 0.95 micron, and uses the nearby 0.75 micron band as an overall reflectivity reference. [37] As regolith ages, it darkens at most wavelengths, 0.75 microns in particular, but the ratio of 0.95 micron to 0.75 micron reflectivity will increase. At infinite age, this ratio approaches about 1.26 and the 0.75 micron reflectivity approaches a few percent. Depending on regolith composition, the approach to this infinite-age point differs, but a sample that is far from this point is reliably young. This distance allows one to rank regolith samples in age. OMAT changes rapidly over the first few million years of exposure. One can test this by looking at the OMAT aging of rays ejected from fresh crater impacts, [38] and the OMAT value is nearly halved in going from 2 million to 50 million year old deposits. Over the course of a billion years OMAT slowly reaches a point were it does not change appreciably.

   Shocks caused by this outgassing are not sufficient to fracture regolith particles, just agitate them. With OMAT we might imagine that volatile stirring of the regolith in various location might be split into several domains. Agitation of the regolith by outgassing might be so rare that OMAT signals volatile release events either fade away or exist as single, isolated features on the lunar surface. The timescale for such fading is of order several hundred million years. This depends in part on how deeply regolith is agitated. Explosive outgassing will mix the entire thickness of regolith, at least several meters thick, so mixing by micrometeorite impacts ("gardening") is completely overwhelmed by the action of gas. If fluidization occurs only at shallow depths as gas approaches the vacuum, gardening effects could win out over volatile mixing over this thin layer in the time it take OMAT to age significantly. For 100 million years this corresponds to about 10 centimeters depth. Another domain occurs if outgassing is more active, and can homogenize the entire depth of regolith into a uniformly OMAT-aged mixture. If every dust grain spends at least about 100 million years within 10 centimeters of the surface, OMAT-aging could be homogenized by outgassing over an entire several meters depth (in the several billion years since the flooding of the maria), so further mixing has little effect on the apparent age of the regolith surface, and that surface would not look unusually young. If the observed optical transients are tied to outgassing, the observed rate of many per year would likely correspond to frequent mixing.

OMAT and outgassing may join in a dramatic circumstance: areas where outgassing from the interior erodes regolith down to the bedrock (or at least the megaregolith underlying the regolith). An example of this is the Ina feature in Mare Imbrium (called a crater, caldera or the "D" feature: Figure 6). Its uniqueness was noted as early as *Apollo 15*, but OMAT showed its surface to be a few million years old or less. (Schultz et al. point out a similar feature in Rima Hyginus, and two smaller ones, all in the maria.) Subsequent high resolution images show the rough, low-lying surface to be almost crater-free and probably young and rocky, whereas the softer, elevated landscape has a cratering density close to that of the surrounding maria. [39] The hypothesis that the regolith is eroded by outgassing from low-lying areas may still be considered the favored hypothesis, but not by all investigators.

There are a few ways to approach changes caused by outgassing, and some are works in progress. Unfortunately LROC-NAC has no 0.95 or 1.8-micron filter (nor did *Lunar Orbiter* or Apollo). *Kaguya-SELENE*'s Multiband Imager had a 0.95 micron filter, but at 20-meter resolution. A way to look for small-scale changes, given the unfortunate constraints on filters, employs broadband reflectivity (as in the 0.75 micron OMAT band). Comparing features at different epochs has little meaning unless illumination and viewing geometries are similar, since reflectivity varies with these passive conditions. We have found one new feature at a TLP site that changed too radically for passive effects alone; [40] we hope for LROC-NAC to recreate the conditions for this site as first imaged by *Lunar Orbiter V*.

Another work in progress is to catch outgassing in action. Not only do we have our model of how explosive outgassing might produce a TLP-like event, but we have the historical record of TLPs themselves to indicate the magnitude and timescale of optical transient that we should search for. Automated telescopes and CCD imagers now make it possible to continuously monitor the Moon's Near Side from Earth, and the presumed timescale of the purported events mean an image needs to be taken perhaps several times per minute. Changes in the Moon over the course of an hour will be negligible due to normal diurnal changes in shadows, except at the terminator itself, and software for analysis of other changes via image differencing algorithms. (See Figure 7.) If we are going to study TLPs, and perhaps even use them as a probe of lunar outgassing, it is important to derive a sample that can be evaluated digitally, objectively and with known observer selection biases. With robotic monitoring new data may overcome the historical database and put the field on a new footing. Our group has collected roughly a quarter million such monitor images of the Near Side and will present their relevance to TLPs at a later time. Monitoring lunar optical transients can have other purposes as well. [41]

To close we should take notice of the Aristarchus Plateau. Obviously the region around Aristarchus is special, with roughly half of all TLP reports, as well as half of the radon-222 events. Figure 8 shows the distribution of these report loci that are well localized (to within 10 kilometers – some just refer to the whole crater, 43 kilometers in diameter). One can see that they are confined to only the southeastern quadrant portion of the Plateau, about 10,000 square kilometers.

The crater itself is one of the youngest for its size on the Moon, 175 million years old, and the Plateau is surrounded to the South, East and West by some of the youngest lunar surfaces, as measured by crater counts, 1 billion years old or less. [42] The Plateau itself is covered with a mantle of pyroclastic deposits from the huge volume of volcanic effusion that is evident in the many volcanic channels, or rilles, evident on the Plateau, flowing off it, and in the surrounding landscape. [43] The largest of these, Vallis Schröteri, is by some definitions the largest rille on the Moon and slices through the Plateau. It also appears associated with a large number of reported optical transients (Figure 8).

    Within Vallis Schröteri are several units; not only is there the large rille itself, up to 10 kilometers across, but an actually longer and more sinuous inner rille that is typically 1 kilometer wide and tends to transition from one wall to the other of the larger rille, having a flat bed a few hundred meters deeper than the floor of the larger rille. The walls of the larger rille are marked by great mass wasting, with boulders typically a few meters but sometime ten times larger rolling onto the inner and outer rille. Figure 9 shows examples of these units, from recent LROC-NAC images. The crater count on these units are highly varied. The Plateau itself has the highest crater count, the outer rille a crater density about one-half that of the Plateau, but the inner rille about one-tenth of the Plateau, making it one of the least cratered lunar surfaces. While the rilles are recessed from the Plateau and therefore subject to a smaller solid angle of incoming impactors, this is only a 10–20% effect. Also, on the Plateau, many features are obviously secondary impacts from the formation of Aristarchus itself. Much of both rilles, however, are aligned radially to Aristarchus crater, including the portions shown in Figure 9, so they should have received a full flux of secondary impacts. Something about the material within the rilles, especially the inner one, either erases or simply cannot sustain cratering. We speculate that this might be one area were the action of lunar volatiles might be apparent, either do to disturbances due to outgassing, modification of the inner rille's bed material, or perhaps in reaction to the Aristarchus impact, at only 175 millions years ago an order of magnitude younger than much of the surrounding Plateau mantle. There is much unsettled about the role of lunar volatiles, and it is appropriate to leave the reader with a mystery.

= = = = = = = = = = = = = =

**TABLE 1: Number of Transient Lunar Phenomenon Reports, by Feature: Raw Report Count from Middlehurst et al. 1968**

Number of
<u>TLP Reports</u>     <u>Feature  (Latitude, Longitude)</u>
 122            Aristarchus (24N 48W)
  40            Plato (51N 9W)
  20            Schroter's Valley (26N 52W)
  18            Alphonsus (13S 3W)
  16            Gassendi (18S 40W)
  13            Ross D (12N 22E)
  12            Mare Crisium (18N 58E)
   6 each      Cobra Head (24N 48W); Copernicus (10N 20W);
             Kepler (8N 38W); Posidonius (32N 30E); Tycho (43S 11W)
   5 each      Eratosthenes (15N 11W); Messier (2N 48E)
   4 each      Grimaldi (6S 68W); Lichtenberg (32N 68W); Mons
             Piton (41N 1W); Picard (15N 55E)
   3 each      Capuanus (34S 26W); Cassini (40N 5E); Eudoxus (44N
             16E); Mons Pico B (46N 9W); Pitatus (30S 13W); Proclus
             (16N 47E); Ptolemaeus (9S 2W); Riccioli (3S 74W);
             Schickard (44S 26E); Theophilus (12S 26E)
   2 each      1.3' S.E. of Plato (47N 3W); Alpetragius (16S 5W);
             Atlas (47N 44E); Bessel (22N 18E); Calippus (39N 11E);
             Helicon (40N 23W); Herodotus (23N 50W); Littrow (21N
             31E); Macrobius (21N 46E); Mare Humorum (24S 39W);
             Mare Tranquilitaties (8N 28E); Mons La Hire (28N 26W);
             Montes Alps, S. of (46N 2E); Montes Teneriffe (47N 13W);
             Pallas (5N 2W); Promontorium Agarum (18N 58E);
             Promontorium Heraclides (14N 66E); South Pole (90S 0E);
             Theaetetus (37N 6E); Timocharis (27N 13W)
   1 each      Agrippa (4N 11E); Anaximander (67N 51W);
             Archimedes (30N 4W); Arzachel (18S 2W); Birt (22S 9W);
             Carlini (34N 24W); Cavendish (24S 54W); Censorinus (0N
             32E); Clavius (58S 14W); Conon (22N 2E); Daniell (35N

31E); Darwin (20S 69W); Dawes (17N 26E); Dionysius (3N 17E); Endymion (54N 56E); Fracastorius (21S 33E); Godin (2N 10E); Hansteen (11S 52W); Hercules (47N 39E); Herschel (6S 2W); Humboldt (27S 80E); Hyginus N (8N 6E); Kant (11S 20E); Kunowsky (3N 32W); Lambert (26N 21W); Langrenus (9S 61E); Leibnitz Mts. (unofficial: 83S 39W); Manilius (15N 9E); Mare Nubium (10S 15W); Mare Serenitatis (28N 18E); Mare Vaporum (13N 3E); Marius (12N 51W); Menelaus (16N 16E); Mersenius (22S 49W); Mont Blanc (45N 0E); Montes Carpatus (15N 25W); Montes Taurus (26N 36E); Peirce A (18N 53E); Philolaus (72N 32W); Plinius (15N 24E); Sabine (1N 20E); S. of Sinus Iridum (45N 32W); Sulpicius Gallus (20N 12E); Taruntius (6N 46E); Thales (62N 50E); Triesnecker (4N 4E); Vitruvius (18N 31E); Walter (33S 0E)

Not counted: 4 (global lunar changes), 14 ("cusp" events), 43 (events w/ unknown coordinates)

/////////////////////////////

## APPENDIX 1

Most TLP reports are visual, and many in recent times originate from amateurs. Before the 20th century, however, many TLP reports came from reputable, professional astronomers, even famous ones: Wilhelm Herschel in 1783-1790 reported TLPs six times (also discovered planet Uranus, several moons, infrared light), Edmond Halley in 1715 (Astronomer Royal of Halley's comet fame), Edward Barnard reported a series of TLPs in 1889-1892 (discovered 17 comets, a moon of Jupiter and showed novae are exploding stars), Ernst Tempel in 1866-1885 (discoverer of 21 comets), Johann Bode in 1788-1792 (famous celestial cartographer), George Airy in 1877 – and confirmed independently (famous Astronomer Royal), Heinrich Olbers in 1821 (established asteroid belt's presence), Johannes Hevelius in 1650 (pioneering lunar topographer), Jean-Dominique Cassini in 1671-1673 (director of l'Observatoire de Paris), Camille Flammarion in 1867-1906 (founded

Société Astronomique de France), William Pickering in 1891-1912 (co-founded Lowell Observatory), Johann Schröter in 1784-1792 (first to notice the phase anomaly of Venus), Friedrich von Struve in 1822 (founded Pulkovo Observatory), Francesco Bianchini in 1685-1725 (measured Earth's axis precession), and Etienne Trouvelot in 1870-1877 (noted astronomical observer).  In the twentieth century noted astronomers reporting TLPs include Dinsmore Alter (in 1937-1959), Zdeněk Kopal (in 1963) and Sir Patrick Moore (in 1948-1967).  Of course, in 1821-1839 Franz von Gruithuisen reported luminous and obscured spots changing on the Moon, yet also wrote about the Moon being inhabited and dotted by cities!  (He was also first to conclude that lunar craters result from meteorite impacts.)

Since the year 1900 and especially since 1957 these reports are made increasingly by amateurs, some of whom are not credible.  Additionally, looking closely at the report catalog one notices observers that produce significantly discrepant results.  For instance James C. Bartlett of Baltimore in 1950 – 1976 reported a record 114 TLPs, 101 of them were "bluish."  There are only 113 reports of blue or violet TLPs, so this category is almost entirely the result of reports by Bartlett.  We do not include the Bartlett events in Figure 2, Table 1 or further analysis.  Nonetheless Bartlett has a reputation as a careful observer and one of his non-bluish reports have been confirmed retroactively as fleeting sunrise illumination of local terrain previously shadowed by a crater rim. [44]

Subjectivity is perilous in these visual reports, and we will consider statistical investigations concerning whether there is consistent and/or physically meaningful information in these reports.  For a moment let us step back and look at a few notable reports and some generalizations about the nature of these data.  Here are some interesting reports:

On November 10 of 557 AD a light reportedly appeared on the thin crescent Moon.  This is the oldest known TLP report (Cameron 1978, Report #1 [45]) and, not surprisingly, is not well documented.

A similar naked-eye report comes from 1668 November (26?) from several people in the Boston area (C78.9; M68.4): a bright spot between the two "horns" of the crescent Moon. Eleven naked-eye reports of bright spots on the Moon at various phases (more crescent) are listed, roughly once per century. This count excludes the few naked-eye reports made by observers *over the Moon* i.e., Apollo astronauts, which we investigate below.

On 1178 June 18 or 19 (Julian date), as chronicled by senior monk Gervase of Canterbury, five men or more reported "a flaming torch sprang up, spewing out, over a considerable distance fire, hot coals and sparks" between the horns of the crescent Moon. [46] This is the object of modern speculation and controversy that has not yet run its course. (C78.5).

In 1650 Johannes Hevelius ("founder of lunar topography") reported the first telescopic TLP, a transient "blood-colored mountain" (*mons porphyrie*) [47] near crater Aristarchus (C78.8, M68.3). In 1671 – 1673 Cassini at l'Observatoire de Paris reported three TLPs that would prove to become fairly typical: a small white cloud, a white spot, a nebulous spot, in crater Pitatus and Mare Crisium (C78.11–13; M68.6–8).

There are some fairly dramatic TLP reports that are *not* typical. A few notable ones include:

Seven times over 1783 – 1790, Wilhelm Herschel reported TLPs, sometimes supported by other (non-independent) observers. In March – April 1787 he reported a complex series of events on the Moon's northwestern quadrant, especially around crater Aristarchus. On March 13, [48] he reported three bright spots appearing on the dark side. (Herschel had reported similar spots during observation of an occultation of a star by the Moon in 1783.) On April 19 he reported a red glow, which he described as a "volcano" near Aristarchus and two other points in the vicinity of craters Menelaus and Manilius on the dark side. On the next night Herschel reported all three points still glowing, with the

Aristarchus site brighter and larger (at least 5 kilometers across), still glowing red. (M68.24–26; C78.31–33)

Over the centuries Herschel's 1787 lunar reports have been dismissed, despite him being one of the greatest observational astronomers. The legitimacy of this attitude may turn with the veracity on several similar reports two centuries later. On 1963 October 30 two cartographers for the U.S. Air Force, James A. Greenacre and Edward Barr, were visually mapping the Moon at the Lowell Observatory 24-inch telescope near Flagstaff, Arizona. They were surprised to see red spots around crater Aristarchus that Greenacre described as similar to "looking into a large, polished gem ruby." This inspired much interest, and a campaign was engaged over the next month to observe Aristarchus. A similar but short-lived event was reported on 1963 November 11, and then a 75-minute event on November 28 was observed not only by Greenacre and Barr, but Flagstaff Observatory director John S. Hall, Clyde Tombaugh (discoverer of Pluto), as well as several observers in Flagstaff and elsewhere in the United States (including Perkins Observatory in Ohio, with confirmation of the reddish spot). There is an independent report from New Jersey, but the timing of this is uncertain. Both the October 30 and November 28 reports correspond roughly to local lunar sunrise. (M68.440–441, 443–445; C78.778,780,783–785) [49]

Adding to the notoriety of the Greenacre & Barr reports were photographs made by astronomers Zdeněk Kopal and Thomas Rackham on 1963 November 1 – 2 and 10 – 11 showing a large-scale enhancement of red light around Aristarchus and two other craters, which was largely confirmed independently by Patrick Moore in the second case. Kopal published his results in *Scientific American*, further increasing public interest. [50] (C78.779, 782; M68.442)

There are at least nine photographic events in Cameron (1978), the earliest from 1953. Many are unpublished, but with dramatic exceptions. The most notable of these are the first TLP photographs made by a professional astronomer. These started the high level of awareness about TLPs that characterizes the later part of the twentieth century and radically changed the characteristics of the report catalog, as we will describe. On 1956 October 26, Dinsmore Alter took a careful sequence of photographs on the Mt. Wilson Observatory 60-inch telescope of craters Alphonsus and Arzachel, in infrared (Kodak I-N photo emulsion) and blue-violet (II-O emulsion) light, allowing differential measurement of the imaging properties in time and wavelength between the two craters. There is a fairly apparent obscuration of the floor of Alphonsus not seen later in time or in Arzachel, and that is apparent in the violet but not infrared as if some scattering cloud is present. [51] A similar effect in crater Purbach was photographed on 1970 April 14 by Osawa but not published. (Cameron 1978) On 1959 January 23, Alter recorded (but never published) a photograph of a bright blue glow on the Aristarchus floor, which then turned white. (M68.381; C78.653)

On 1953 November 15 radiologist and amateur astronomer Leon Stuart was observing the Moon when he noticed a bright flash near its center and managed to photograph it. (Figure A.1) He did not observe the start nor end of the flash, but estimated its duration between 8 and 30 seconds (although son Jerry Stuart is quoted as saying that 8 seconds is probably too short a time to take the photo.) Fifty years later planetary scientist Buratti and student Johnson noted that the flash's location on Stuart's image corresponds to a fresh, 1.5-kilometer diameter crater and proposed that this was the result of an impact seen by Stuart. Later objections to this hypothesis are several: more careful measurement of the positions of the crater and Stuart's flash image show they do not coincide, the crater was found on pre-1953 photographic plates, and the age of the crater's surface was shown to be too old. Furthermore, theoretical arguments show that a crater this size would cause a plasma fireball visible for only about 1 second, not 8 – 30. Stuart's 1953 event appears real, but did not result from an impact. [52] (M68.312, C78.559)

Two unpublished photographs are claimed to show red spots in craters Aristarchus and Maskelyne, respectively, with the latter report apparently confirmed by separate visual observers. Two other unpublished photographs involve brightenings of Aristarchus. (C78.876, 1145)

More recently Cameron (1991) presents fairly dramatic photographs of a glowing, reddish-gray patch moving on the floor of crater Piticus, as observed by Gary Slayton of Fort Lauderdale, Florida on 1981 September 5. See Figure A.2. (C06.152)

During a polarimetric program at l'Observatoire de Paris for lunar surface texture analysis, Dollfus recorded on 1992 December 30 a brightening in the center of crater Langrenus, and with it an associated increase in the degree of polarization. [53] He favored the interpretation of scattered light from suspended dust, presumably due to gas release. The phenomenon did not change over six minutes but had decreased when observed three days later. Similar polarimetric changes interpreted as dust above the crater were recorded on at least two occasions in Aristarchus, but the timescale is unclear. (C78.820)

Other kinds of permanent records of TLP have been made, including spectroscopy. Most notably, Nikolai Aleksandrovich Kozyrev was intrigued by Alter's photographs of transient obscurations in 1956. On 1958 November 3 at Pulkova Observatory in the Crimea while observing the crater Alphonsus in preparation for taking spectra, he noticed the crater's central peak turning red and varying in brightness. He began a series of photographic spectra including the crater's central peak and much of its diameter, some recording a series of absorption bands that show similarity of those due to molecular carbon ($C_3$). Kozyrev saw another TLP on 1959 October 23 in Alphonsus and took another spectrum showing bands in emission which he interpreted as $C_2$ and $C_3$. The molecular carbon explanation was difficult to accept for many and Kozyrev's reputation suffered. The incident marked a substantial rift between the Soviet and American lunar science community that had long-term effects. [54] (C78.703,723; M68.409,423)

Several reports by simultaneous but geographically distant observers of the same events on the lunar surface are recorded e.g., 1895 May 2, for 12 – 14 minutes on the floor of crater Plato, Leo Brenner reported a streak of light, while independently Philip Fauth reported bright, parallel bands. [55] Plato will be a topic of interest shortly.

A singular TLP, naked-eye *and* telescopic, reported simultaneously by independent observers 385,500 kilometers apart, occurred on 1969 July 19, just before the first human lunar landing. It was reported by two astronomers in West Germany and two of the three *Apollo 11* astronauts. (The mission transcript does not indicate Neil Armstrong as witness.) Buzz Aldrin and Mike Collins reported a strange darkside surface appearance near Aristarchus during a 1 – 2 minute period in which ground-based observers saw a similar phenomenon at likely the same location. This report is special in that discussion around the time were recorded verbatim, so one can understand some of the human factors involved with the report, unlike many in the historical sample. [56] The crew had been alerted to recent activity near Aristarchus, so this was not fully independent. Aldrin dismissed the phenomenon as being explained by anomalous backscattering (which you can observe on an airplane by noticing the bright spot surrounding your plane's shadow), but the geometry was not correct for this effect. [57] (C78.1165)

Three instances of rapid, bright flashes apparently from the lunar surface were observed on *Apollo 16* (by Ken Mattingly) and *Apollo 17* (by Jack Schmitt, Ron Evans – *not* Cernan per Cameron). I will not analyze these statistically, but they are worth mention. They are documented in the mission transcripts, debriefings, preliminary science reports and Cameron's catalog. The two *Apollo 17* reports were tied to the flooded crater Grimaldi and Mare Orientale, respectively. The first locus, and even the second, while indistinct, are sites of some of the few outgassing events detected by non-TLP means (both during *Apollo 15* – see below). Grimaldi is a persistent TLP site; Orientale is too close to the limb to be seen well from Earth. This is interesting because few if any of flashes were seen *not* coming from the direction of the lunar surface, so their cause being cosmic ray interactions with the retina or vitrious humour of the eye is problematic. The *Apollo 16* event is hard to localize being on the dark side and purely visual. Being on the far side, it cannot be tied explicitly to other TLPs. Nonetheless, Mattingly saw it as coming from below the horizon and therefore ostensibly the lunar surface. As best as I can reconstruct from his description, Mattingly was looking in the vicinity of crater Korolev on the far side.  This is highly uncertain.  (C78.1331,1352,1354)

I have discussed these events with Schmitt and Mattingly.  (Evans died in 1990.)  Both were well aware of cosmic ray-induced flashes within the human eye.  Schmitt is nearly certain this is not what he saw; Mattingly is somewhat less sure because *he never saw a cosmic ray flash* otherwise, but describes the event as completely point-like and instantaneous, which varies with documented description of cosmic ray events.  Schmitt also describes his event as point-like and instantaneous, although there is slightly contrary reference in the *Apollo 17* transcript (within one minute of the report – "Schmitt: It was a bright little flash right out there near that crater.  See the crater right at the edge of Grimaldi.  Then there is another one north of it.  Fairly sharp one north of it is where there was just a thin streak of light.")  Rather than dwell on the memory of four-decade old events, one can make a brief inquiry into whether these events are easily explained by meteorite impacts on the lunar surface.  Quantitatively, this seems likely for two of three cases. [58]

These many anecdotal TLP reports are a rat's nest of selection biases and possible observer errors.  Furthermore, I have described above many of the more dramatic ones; most are small spots.  Rather than dwelling on individual events, I will explain below how we might be able to glean insight from these many reports statistically.

A total of 71 reports in the Middlehurst et al. 1968 catalog include duration estimates (to a range smaller than a factor of 2, which excludes for instance the Stuart 1953 TLP).  This is not a statistical sample, but the reports indicate prolonged occurrences; binned in $\sqrt{10}$ intervals from 60s to 19000s (with the longest event being 18000s and the shortest 60s) the duration distribution is: 60–190 seconds, 7 reports; 191–600s: 9; 601–1900s, 27, 1901–6000s, 23; more than 6000s: 5.  These effects are not so rapid as to not allow re-inspection (albeit by the same observer in most cases).  The median length is 15 minutes, but this is almost certainly an overestimate because shorter events are easier to miss.

Assuming this is a linear correction, the median duration is closer to 3 minutes. Nonetheless, even during longer observations, internal changes are often seen on rapid timescales. A few examples:

1964 May 18, Universal Time 03:55 – 05:00: Southeast of crater Ross D; 65 minutes long – "White obscuration moved 20 mph, decreased in extent. Phenomenon repeated." (C78.818, M68.458)

1966 April 12, UT 01:05: crater Gassendi, 18 minutes long – "Abrupt flash of red settling immediately to point of red haze near NW wall. Continuous until 01:23." (C78.925, M68.536)

1966 September 2, UT 03:16 – 04:18: crater Alphonsus, 62 minutes long – "A series of weak glows; Final flash observed at 04:18" (C78.971, M68.549)

There are four cases in Middlehurst et al. 1968 described as sudden, isolated flashes of light, and these are not correlated with meteor showers (occurring on 1945 October 19, 1955 April 24, 1957 October 12, and 1967 September 11). None of these are well-placed with respect to known meteor showers. (April 23 is the peak of the Pi Puppid meteor shower, but these are strong only near the perihelion of comet 26P/Grigg-Skjellerup, which occurred in 1952 and 1957, not 1955.)

Several studies have been made of meteoroid impacts on the Moon by patrolling with video cameras on Earth. (A meteoroid is a small object in solar orbit that would become a meteor if it entered Earth's atmosphere.) This is most easily accomplished during meteor showers, during which impact rates can be thousands of times average. One problem is that video cameras produce artificial flashes that might be mistaken for impacts, but this has been avoided by requiring coincidence between simultaneous monitors. [59] These impacts are somewhat by necessity small and short, passing primarily in 0.02 seconds. By covering more area and time, one can also wait for occasional, kilogram-range impacts, which make 0.1 second duration events. The most impacts have been detected by Marshall Space Flight Center's monitoring program, at least 240 thus far, the longest lasting 0.5 second. [60] Shower meteoroid impacts are not uniform, but tend towards lunar location where the radiant point for the meteor shower is overhead. Nonetheless, TLPs are much more restricted in location than meteoroid impacts. For a 1-second event, thousands of tonnes in impactor is needed, leaving a noticeable crater about 200 meters across. Even comets striking Jupiter e.g., Shoemaker-Levi 9, with masses of billion of tonnes produced flashes of only 5 – 10 seconds. Few TLPs are explained by meteoroid impacts on the Moon.

Those many impacts affect the lunar atmosphere: they make it glow like a sodium vapor lamp. (See Figure A.3) Meteoroid impacts erode many thousands of tonnes of lunar regolith per year. [61] Sodium is lost in this process and also by "desorption" – vaporization in this case by heat or solar ultraviolet irradiation, or "sputtering" – loss from particle collisions, in this case by the solar wind. Scientists are still undecided regarding which processes dominate. [62] A similar process occurs for potassium, and searches have been made for several other elements. [63] Solar radiation is reprocessed efficiently by these atoms' resonant scattering transitions (5893Å for sodium, 7699Å for potassium). Only a tiny amount of these atoms is found (rarely more than 100 per cubic centimeter), so they contribute negligibly to the lunar atmospheric mass. Nonetheless, they surround the Moon in a visible plume to several dozen lunar radii, easily confused with the Moon's dust halo if one does not separate their signals spectroscopically.

Are optical transients obviously correlated with physical processes on the Moon? Several teams of investigators have correlated existing TLP lists, especially the timing of the reports, with external processes. This is a highly uncertain procedure, since when TLPs are reported depends not only on when hypothetical physical processes on the Moon might occur, but also when humans are observing. Studies have looked for correlation of TLP activity with maximum stress from Earth's tidal force but largely failed using large report samples. [64] Early on there was some hint of such a correlation for reports associated with crater Aristarchus only, which later studies do not rule out. [65] Middlehurst's 1977 study also claims a close pairing of TLP and deep moonquake loci, with epicenters always within 150 kilometers of a TLP site.

Are TLPs real? These may be interesting phenomena tied to meaningful physics at and below the Moon's surface. We would be mistaken to dismiss the entire topic as optical illusion or figments of weak-minded imagination without some scientific evaluation of its reality. However, existing evidence resides in a horrendous hairball of a dataset. We could imagine some way of taking a more controlled and regularly sampled set of observations, either from Earth's surface or the Moon, but this would involve significant expense. Before that effort we should ask if it is possible to glean some meaning despite the observer biases, errors and unknown selection criteria. Let us try.

Which TLPs are real? Rather than trying to decide which individual observers are reliable, and which TLP sites are favored because of the habits and interests of observers, we can test which locations on the Moon indicate transient activity regardless of the observer. To understand the spatial distribution of TLPs, we must deal statistically with the huge selection effects introduced by observers, most of whom never intended their reports for a statistical database. This task is not modest; we need to calibrate all of the observations of the Moon made (or not made) by all of the human eyeballs over several centuries. How do we possibly deal with historical and even psychological issues that these effects imply, as well as the physical/mathematical ones? This is the major burden; nonetheless, there are regularities that we can exploit. Before I describe these tests and their results, let us take a moment to consider how lunar observing has changed with time.

Many works review the history of selenography: there were systematic naked-eye lunar observations starting in Europe in the 1400 – 1500s, and much earlier in China (but I find no early Chinese reports of TLPs). Observations increased greatly in intensity in the mid-17th century after the invention of the telescope, and early in the next century it was appreciated that libration effects required more observation of the lunar limb. Mapping the Moon increased in detail, with lunar coordinates adopted by the mid-to-late 18th century. Lunar cartography and topography become forefront science. From late-1700s to early 1900s, visual mapping of the entire Near Side was active, with some concentration on terminator regions to better sense features' relative elevations. By the early 20th century, visual observation increasingly lost out to photography, suppressing sensitivity to TLPs due to decreased sampling frequency ("cadence") of photographic plates versus the eye, and lost color information. TLP reports by professional astronomers began to die out for this reason, I speculate. This break in TLP reports appears both the 1978 Cameron and 1968 Middlehurst catalogs; indeed, the 1927-1931 gap divides the Middlehurst catalog at its median report epoch (its half-way point). For post-1930 reports, two-thirds come after 1955, after which TLPs become commonly known and observing patterns change, as described below. I will concentrate on the pre-1956 and particularly the pre-1930 period, for reasons that will become clear.

Prior to 1956 there was little general concept of what today we call a TLP, perhaps with one giant exception: the crater Linné. It was first noted about 1665 by Giovanni Battista Riccioli. Today we know it is 2.4 kilometers in diameter, young and exceptionally high in reflectivity. Experienced selenographer (Johann Friedrich) Julius Schmidt in 1866 claimed that Linné had disappeared. He and other observers claimed only a protrusion remained. [66] Then in 1867 Schmidt claimed it had grown into a mountain. This began a decades-long debate, but convinced few lunar scientists that a change had occurred. Modern data shows that Linné is a simple crater, bearing no evidence of any historical alteration. In truth, this would not have been called a TLP even if it had changed, given the permanence of the claimed effect. Until 1956 there was little discussion of any sufficiently rapid activity on the Moon.

There is a distinct exception to this seeming inattention. Schröter's 1791 *Selenotopographische Fragmente* treats lunar transients explicitly, citing 16 of his own TLP reports, another by Von Brühl (from 1787), and the 1787 sequence of Herschel, which seems to have peaked his interest.

Schröter had observed TLPs before this sequence but apparently paid more attention to them after hearing of Herschel's. This book was influential with some astronomers in the following century. (We are told that the book in 1840 inspired the teenage Julius Schmidt, later the noted selenographer who initiated the crater Linné controversy, to make a career of the Moon. [67]) The book's concentration on what we now call TLPs (recall that it is subtitled "the lunar surface: its changes and sustained atmosphere") was at first accepted, [68] but within a century thrown into disrepute. [69] While it was cited occasionally throughout the 19th century, beyond Linné (and a reference or two to spots in Plato) its TLP aspects were rarely discussed.

How much were observers aware if each other's TLP reports throughout history? This is important for at least two reasons: if attention is drawn to a particular lunar feature, other reports of that feature's activity are correlated, which must be taken into account in any estimate of statistical significance. Secondly, a publicized report might result in a "hysteria signal" in which inexperienced observers are drawn in and begin to make erroneous reports, or even experienced observers might forego their usual caution. Phenomena such as Earth's atmospheric refraction or "seeing" (scintillation, "twinkling"), internal reflections within telescopes, and other non-lunar effects can be mistaken for TLPs by inexperienced and/or overly eager observers.

One way to show how much observers paid attention to each other's TLP reports are via counts of publication citation cross-referencing. The Astrophysics Data System by NASA and the Smithsonian Astrophysical Observatory covers astronomical and planetary research publications relevant to TLP reports as far back as the 17th century, [70] and tracks citations and references for these articles. (ADS's citation/referencing service becomes less reliable for articles stretching beyond mid-19th century. It still covers about 93% of TLP reports.) To summarize the results from using ADS, [71] after splitting the sample into three intervals: before 1930 (with 1930 being the dip in TLP counts), 1930 – 1955 (with 1956 being the start of the Alter – Kozyrev TLP era), and 1956 and later (to the end of the 1968 Middlehurst catalog), one sees a dramatic shift after 1956: TLP reporters tend to be largely oblivious to each other and the existence of reputed transient lunar phenomenon associated with the lunar feature they are studying or any other, until 1956, after which nearly all TLP observers refer to TLPs, and many of which are observing the Moon in search of TLPs.

The few articles before 1956 that do mention changes on the Moon (postdating Schröter's work in 1791) only strengthen the point: they do not relate to what we would call TLPs, and seem ignorant of Aristarchus. The only pre-1956 citation involving both "change" and "aristarchus" is Haas (1938) referring to periodic changes appearance of the inner eastern wall of Aristarchus over nine-day intervals, hence not a TLP. The two other Aristarchus citations concern the same subject and seem in reaction to Haas (1938). [72] This kind of statistical correlation between events we would guard against *if* they involved TLPs. Furthermore, Birt (1870) and Whitley (1870) provide a historical overview (1787–1880) of visual observations of Aristarchus (and Herodotus) while conducting a spirited debate about the nature of features including possible changes in their appearance. They mention small, possible changes, but give them no special significance, nor mention anything that today we might call a TLP. [73]

By 1913, however, there appears to be some scholarly awareness of reported activity at Aristarchus; witness the summary by Maunder of reports by Herschel that he "thought he was watching a lunar volcano in eruption," and by Molesworth and Goodacre who "each on more than one occasion, observed what seemed to be a faint bluish mist on the inner slope of the east wall... for a short time. Other selenographers too, on rare occasions, have made observations accordant with these, relating to various regions on the Moon." [74] Maunder balances this with skepticism e.g., "one of the most industrious of the present-day observers of the Moon, M. Philip Fauth, declares that as a student of the Moon for the last twenty years, and as probably one of the few living investigators who have kept in practical touch with the results of selenography, he is bound to express his conviction that no eye has ever seen a physical change in the plastic features of the Moon's surface."

Another telling summary from this period by W.H. Pickering (1892) asks "Are there present Active Volcanoes upon the Moon?" and does a quantitative study of candidate volcanoes on the Moon, merging two lists with a total of 67 craters, and then discussing in turn many of the 32 craters common to both lists. Most of these are then eliminated for various reasons, then he starts discussing the rest in turn.  Aristarchus is mentioned, but dismissed as non-volcanic, while some features are taken much more seriously (Bessel, Linné and Plato).  There is *no* discussion of activity that we would call TLPs, only the long-timescale or permanent changes.  Pickering reported several TLP shortly before and then after this paper --- statistically marginal in themselves, 14 in our sample, and 9% of Aristarchus and vicinity --- but I see no evidence that these induced a spate of further Aristarchus reports, at least until publication of his book in 1903, [75] and probably not even then.  This was late in the pre-1930 period, regardless.  A statement made by Elgers in 1884, reviewing Aristarchus, Herodotus and the Plateau, does not mention anything like TLPs but makes a telling statement:  "Although no part of the moon's visible surface has been more frequently scrutinized by observers than the rugged and very interesting region which includes these these beautiful objects, selenographers can only give an incomplete and unsatisfactory account of it…" [76] Aristarchus again received close scrutiny in 1911 with Robert Wood indicating that it might contain high concentrations of sulfur, but this did not produce a spate of Aristarchus TLP reports.  Indeed, Wood discusses volcanism in the context of Aristarchus (sometimes known "Wood's Spot") and seems unaware of the number of TLP reports in the vicinity. [77]

That is sufficient history for now; cognizant of it, let us try to make use of the catalog of TLP reports in a way that minimizes some of these observer effects that we have described.  We want to eliminate the effect of a class of discrepant reports in a portion of the catalog by comparing them to another portion absent that effect.  One obvious way to reject discrepant effects is by use of the median of a distribution of supposedly consistent values, which is a much more robust way of estimating the average than a simple mean.  If fewer than half of the values are truly consistent and not discrepant, taking the median will return an average value chosen from the consistent values.  Here the "value" will be the number of TLP reports confined to a specific area of the lunar surface.  Specifically, we bin the counts seen in Figure 2 into square "pixels" 300 kilometers on a side.  Furthermore we subdivide these report counts to a greater extent when possible. [78] We split all of the reports that correspond to each pixel into three groups according to one of several criteria *that reflect differing properties of the observers* and presumably not conditions related to the Moon, then take the median of these three counts for each pixel.  (For convenience, we label each pixel with the corresponding name from the features listed in Table 1.)  For instance, we can split the Middlehurst catalog into three equal portions by splitting the historical timeline at 1892 and 1956.  The resulting median counts are shown in Table A.5: the Aristarchus Plateau (including Schröter's Valley, Cobra Head & Herodotus) contribute 43% of the counts, and Plato 12%, followed by a smattering of features each comprising no more than 3%.  If we reject every report after 1955 and divide the remainder into three, the results (Table A.6) are nearly identical: Aristarchus Plateau 38%, Plato 16%, Mare Crisium 7% and Tycho 5%.  (Mare Crisium and Tycho contributed 2% and 1% in the previous median.)  One can reject discrepant counts with only *two* values if one assumes that the only deviations are positive-going.  In his assumption the robust estimator (taking the place of the median) is just the lower of the two counts.  By splitting the sample in two at 1930, one finds similar counts again (Table A.7): Aristarchus Plateau 64%, Plato 15%, and a number of feature less than 3% apiece.

What results if we split time in a way not sensitive to historical trends, but two season?  Splitting the year into four-month intervals, one finds a median contribution (Table A1.8): Aristarchus Plateau 38%, Plato 16%, Mare Crisium 7%, similar to the historical sample segregation.  Finally, one can split the sample in yet another observer variable other than time which should have little bearing on the physical processes on the lunar surface: where the observer was located on Earth (Great Britain versus continental western Europe versus the rest of the world: primarily the Americas with some reports from Asia and Russia). The median of counts per feature "by continent" (Table A1.9) is Aristarchus Plateau 43%, Plato 17%, Mare Crisium 7% and Tycho 5%. These results are remarkably consistent. For the mean of these five various estimates of the number of reports per feature, one finds (Table A1.10): Aristarchus Plateau 46.7%±3.3%, Plato 15.6%±1.9%, Mare Crisium 4.1%±1.0%, 2.8%±0.8%, Kepler 2.1%±0.7%, Grimaldi 1.6%±0.6%, and Copernicus 1.4%±0.6%.

**TABLE A.1: TLP-related Terms Correlated with "Aristarchus" and "Moon" before 1930**

| Search Term | Number of Citations | "moon/lunar" Cross-Citations | "aristarchus" Cross-Citations |
|---|---|---|---|
| "volcano" | 38 | 14 | 0 |
| "gas" | 1194 | 0 | 0 |
| "atmosphere" | 599 | 22 | 0 |
| "eruption" | 52 | 1 | 0 |
| "flash" | 89 | 0 | 0 |
| "cloud" | 192 | 3 | 0 |
| "nebulosity" | 45 | 0 | 0 |
| "mist" | 2 | 1 | 0 |
| "geyser" | 3 | 0 | 0 |
| "vapo(u)r" | 514 | 0 | 0 |
| "transient" | 9 | 0 | 0 |
| "change" | 893 | 38 | 0 |
| No First Term | – | 2930 | 7 |

**TABLE A.2: TLP-related Terms Correlated with "Aristarchus" and "Moon," 1930 – 1955**

| Search Term | Number of Citations | "moon/lunar" Cross-Citations | "aristarchus" Cross-Citations |
|---|---|---|---|
| "volcano" | 29 | 6 | 0 |
| "gas" | 2214 | 3 | 0 |
| "atmosphere" | 2234 | 47 | 0 |
| "eruption" | 119 | 2 | 0 |
| "flash" | 99 | 3 | 0 |
| "cloud" | 1036 | 9 | 0 |
| "nebulosity" | 81 | 1 | 0 |
| "mist" | 5 | 0 | 0 |
| "geyser" | 1 | 0 | 0 |
| "vapo(u)r" | 698 | 2 | 0 |
| "transient" | 112 | 1 | 0 |
| "change" | 2050 | 26 | 1 |
| No First Term | – | 1124 | 3 |

**TABLE A.3: TLP-related Terms Correlated with "Aristarchus" and "Moon" after 1955**

| Search Term | Number of Citations | "moon/lunar" Cross-Citations | "aristarchus" Cross-Citations |
|---|---|---|---|
| "volcano" | 196 | 54 | 4 |
| "gas" | 5881 | 24 | 1 |
| "atmosphere" | 4797 | 47 | 0 |
| "eruption" | 120 | 7 | 1 |
| "flash" | 262 | 1 | 0 |
| "cloud" | 1850 | 20 | 0 |
| "nebulosity" | 80 | 1 | 0 |
| "mist" | 5 | 0 | 0 |
| "geyser" | 5 | 0 | 0 |
| "vapo(u)r" | 1154 | 8 | 0 |
| "transient" | 458 | 15 | 3 |
| "change" | 4836 | 61 | 1 |
| No First Term | – | 2440 | 10 |

# TABLE A.4: Number of TLP Reports, by Feature: Median of Counts from 1892–1955, Before & After

| Number of TLP Reports | Feature |
|---|---|
| 46 | Aristarchus/Schröter's Valley |
| 13 | Plato |
| 3 | Kepler |
| 2 each | Alphonsus, Eudoxus, Grimaldi, Mare Crisium, Posidonius |
| 1 each | Alpetragius, Peak S. of Alps, Bessel, Calippus, Cassini, Carlini, Copernicus, Daniell, Gassendi, Godin, Hercules, Kant, La Hire, Littrow, Manilius, Mare Humorum, Messier, E. of Picard, SW |

of Pico, Proclus, Promontorium Heraclides, Ptolemaus, Riccioli, South Pole, Tycho

108 total

**TABLE A.5: Number of TLP Reports, by Feature: Median of Counts from before 1956, Split into Three Intervals**

Number of
TLP Reports   Feature

 29            Aristarchus/Schröter's Valley
 12            Plato
  5            Mare Crisium
  4            Tycho
  2            Kepler
  1 each       Alphonsus, Peak S. of Alps, Bessel, Calippus, Cassini, Copernicus, Eudoxus, Gassendi, Godin, Grimaldi, Kant, Lichtenberg, Taurus Mountains, Macrobius, Messier, Picard, Pico, Posidonius, Proclus, Promontorium Heraclides, Ptolemaus, Riccioli, South Pole, Theaetetus

76 total

**TABLE A.6: Number of TLP Reports per Feature: Adopting Lesser of Pre- and Post-1930 Counts**

Number of
TLP Reports   Feature

 66            Aristarchus/Schröter's Valley
 15            Plato
  2 each       Grimaldi, Messier
  1 each       Alphonsus, Bessel, Cassini, Copernicus, Gassendi, Kepler, Lichtenberg, Littrow, Mare Humorum, Mons Pico, Pallas, Picard, Ptolemaeus, Riccioli, South Pole, Theaetetus, Tycho

103 total

**TABLE A.7: Number of TLP Reports per Feature: Median of Counts over Season (January-April, May-August versus September-December)**

Number of
TLP Reports   Feature

 25            Aristarchus/Schröter's Valley

13          Plato
  5          Mare Crisium
  2 each     Copernicus, Eratosthenes, Kepler, Tycho
  1 each     Atlas, Bessel, Cassini, Grimaldi, Hansteen, Helicon, Herschel, Humboldt,
       Hyginus, Kant, La Hire, Lichtenberg, Messier, Picard, Pickering, Pierce A, Pico,
       Posidonius, Proclus, Promontorium Heraclides, Ptolemaeus, Riccioli
73 total

**TABLE A.8: Number of TLP Reports per Feature: Median of Counts over Geographical Region of Observer Location**

Number of
TLP Reports    Feature

 37            Aristarchus/Schröter's Valley
 15            Plato
  6            Mare Crisium
  4            Tycho
  2            Eratosthenes
  1 each       Alphonsus, Atlas, Bessel, Calippus, Cassini, Copernicus, Gassendi, Godin,
       Grimaldi, Hercules, Kant, Kepler, La Hire, Lichtenberg, Macrobius, SW of Pico,
       Posidonius, Proclus, Promontorium Heraclides, Ptolemaeus, Riccioli, South Pole,
       Theaetetus
87 total

# TABLE A.9: Relative Frequency of Robust TLP Report Counts by Feature, Averaging Different Methods (from Tables A.4 – 8)

Relative
Frequency
± 1 $\sigma$ Error     Feature

46.7%±3.3%      Aristarchus/Schroter's Valley
15.6%±1.9%      Plato
 4.1%±1.0%      Mare Crisium
 2.8%±0.8%      Tycho
 2.1%±0.7%      Kepler
 1.6%±0.6%      Grimaldi
 1.4%±0.6%      Copernicus

 (6.2%±1.2%  *sum of Tycho, Kepler & Copernicus*)
 1.1%±0.5%  Alphonsus, Bessel, Cassini, Messier, Ptolemaus, or Riccioli
 0.9%±0.5%  Eratosthenes, Gassendi, Kant, Lichtenberg, (E. of) Picard, (SW of) Pico (B), Posidonius, Proclus, Promontorium Heraclides, or South Pole
 0.7%±0.4%  Calippus, Eudoxus, Godin, La Hire, or Theaetetus
 0.5%±0.3%  Peak S. of Alps, Atlas, Hercules, Littrow, Macrobius, or Mare Humorum
 0.2%±0.2%  Alpetragius, Carlini, Daniell, Hansteen, Helicon, Herschel, Humboldt, Hyginus, Manilius, Pallas, Pickering, Pierce A, or Taurus Mountains

   ///////////////////////////

## FOOTNOTES:

[1] See, for instance, "Physical Changes upon the Surface of the Moon" by Edmund Neison, 1877, *Quarterly Journal of Science*, 7, 1: "The present condition of the surface of the moon is one of the most interesting and important questions within the whole range of Astronomy… The question of the present condition of the surface of the moon has engaged, therefore, the attention of some of the most eminent astronomers… It is a remarkable circumstance, in relation to this question, that whereas those astronomers who have devoted much time and labour to the study of the moon's surface, and to whom astronomers in general are mainly indebted for our present knowledge of the surface of our satellite, hold in general one view as to the present condition of the lunar surface, astronomers as a body hold a very different opinion. To take a striking instance, scarcely any astronomer known to have devoted time to the study of selenography doubts that many processes of actual lunar change are in progress, and it is doubtful if there is one who could not promptly instance one or more such cases. Yet the general opinion of astronomers appears to be against any such physical changes having occurred. And another instance, almost as striking, exists in connection with the subject of the lunar atmosphere: whilst all selenographers appear to have detected instances where the existence of this atmosphere is revealed, astronomers in general appear to question almost the possibility of its existence, and this in face of the absence of any evidence whatever that there is no atmosphere of the nature supposed.

 It would be an interesting inquiry to ascertain in what manner arose this direct conflict of opinion on this subject, between those who have systematically studied the appearances presented by the moon, and those who have not in the same systematic and assiduous manner examined the lunar surface. It appears to have originated in a short summary by Mädler of the appearances presented by the moon, wherein the differences between the condition of the moon and earth were forcibly stated, and where he pointed out the impossibility of the view that was held in the time of the earlier astronomers, that the moon might be a mere copy of the earth, containing a dense atmosphere, large oceans, abundant vegetation, and animal life, or even human inhabitants. In a condensed and more unqualified form these remarks crept into all astronomical textbooks, and were the main basis of the views commonly held by astronomers."

Neison quotes textbooks of the day calling the Moon an "airless, waterless, lifeless, volcanic desert." Johann Heinrich von Mädler did his lunar research in the 1830s.

[2] For instance: "It has been a question which has been long debated among astronomers, whether the Moon has an atmosphere or not, and, as far as I have been able to learn from reading and verbal inquiry, the question is yet undecided. The best astronomers I have talked with about this matter have told me they could never discover any atmosphere about the Moon." ("Certain Reasons for a Lunar Atmosphere" by Samuel Dunn, 1761, *Philosophical Transactions*, 52, 578)

[3] "The Lunar Atmosphere and The Recent Occultation of Jupiter" by William H. Pickering, 1892, *Astronomy & Astrophysics, Carleton College Goodsell Observatory*, 11, 778.

[4] This is contentious. Charged particle density probably varies greatly with time; nonetheless various measurements seem to conflict. An early result ("Measurement of the lunar ionosphere by occultation of the *Pioneer 7* spacecraft" by J.C. Pomalaza-Díaz, 1967, in *Scientific Report SU-SEL-67-095*, Stanford Electronics Lab) puts an upper limit of 40 electrons per cubic centimeter vs. $1.45\times10^{16}$ cm$^{-3}$ for neutral molecules at Earth's surface. Two other occultation measurements, on *Luna 19* and *22*, provide detections of several hundred electrons per cm$^3$ ("Preliminary Results of Circumlunar Plasma Investigations" by M.B. Vasylyev et al. 1973, *Doklady Akademii Nauk SSSR*, 212, 67; and "Preliminary Results of Circumlunar Plasma Research by the *Luna 22* Spacecraft" by A.S. Vyshlov, 1975, *Space Research*, 16, 945). Recent results agree better with *Pioneer 7* results, again, for charged species. ("Studying the Lunar Ionosphere with *SELENE* Radio Science Experiment" by Takeshi Imamura, 2010, *Space Science Reviews*, 154, 305)

[5] "Molecular Gas Species in the Lunar Atmosphere" by J.H. Hoffman & R.R. Hodges, 1975, *Moon*, 14, 159.

[6] "Lunar atmosphere measurements" by Francis S. Johnson, James M. Carroll & Dallas E. Evans, 1972, *Proceedings of 3rd Lunar Science Conference*, 3, 2231.

[7] One can show that atmospheric density $n$ varies with temperature $T$ such that $n \propto T^{-5/2}$ *except for sticking effects*, which reduce density when $T$ is small (when $T^{-5/2}$ would be large, otherwise).

[8] "Lunar Theory & Processes: Post-sunset Horizon 'Afterglow'" by D.E. Gault, J.B. Adams, R.J. Collins, G.P. Kuiper, J.A. O'Keefe, R.A. Phinney, E.M. Shoemaker, 1970, *Icarus*. 12, 230

[9] "*Surveyor I* Observations of the Solar Corona" by Robert H. Norton, James E. Gunn, W.C. Livingston, G.A. Newkirk & H. Zirin, 1967, *Journal of Geophysical Research*, 72, 815. The

sulfur dioxide, chlorine, hydrocarbons up to octane, krypton, but not xenon or radon. Unintentional contamination from tungsten in the instrument was used to maintain calibration.

[19] "Helium & Hydrogen in the Lunar Atmosphere" by R.R. Hodges, Jr., 1973, *Journal of Geophysical Research*, 78, 8055;

[20] "Lunar atmospheric composition results from *Apollo 17*" by J.H. Hoffman, R.R. Hodges, Jr., F.S. Johnson & D.E. Evans, 1973, *Proceedings of 4th Lunar Science Conference*, 3, 2865.

[21] "Molecular Gas Species in the Lunar Atmosphere" by J.H. Hoffman & R.R. Hodges, Jr., 1975, *Moon*, 14, 159.

[22] *Apollo 12* and *15* ALSEP contained the Solar Wind Spectrometer (SWS), which collected (in seven directional horns) high-energy charged particles (electrons of 6–1330 eV, ions of 18–9780 eV), which originate overwhelmingly in the solar wind. The Charged Particle Lunar Environment Experiment (CPLEE) on *Apollo 14* studied both ions and electrons with energies in the range 50–50,000 eV. Lunar subsatellites launched on *Apollo 15* and *16* also carried a charged particle detector, and the SIM bay boom on *Apollo 15* and *16* carried the Orbital Mass Spectrometer Experiment (OSME) for neutral species, as detailed below. "Molecular Gas Species in the Lunar Atmosphere" by J.H. Hoffman & R.R. Hodges, Jr., 1975, *Moon*, 14, 159.

[23] "Formation of the Lunar Atmosphere" by R.R. Hodges, Jr., 1975, *Moon*, 14, 139

[24] Of course, half-life is the time required for 50% of the substance to decay. $^{40}$K atoms decay into $^{40}$Ar 10.72% of the time and 89.28% into calcium ($^{40}$Ca). About 0.01% of potassium is $^{40}$K. See "Implications of atmospheric $^{40}$Ar escape on the interior structure of the Moon" by R.R. Hodges, Jr. & J.H. Hoffman, 1975, *Lunar & Planetary Science Conference*, 6, 3039.

[25] Apollo LACE results indicate a gas leakage rate from the Moon of $10^{23}$ s$^{-1}$ in $^4$He and $2 \times 10^{21}$ s$^{-1}$ in $^{40}$Ar: "Formation of the Lunar Atmosphere" by R.R. Hodges, Jr., 1975, *Moon*, 14, 139

[26] "Detection of Radon Emanation from the Crater Aristarchus by the *Apollo 15* Alpha Particle Spectrometer" by Paul Gorenstein & Paul Bjorkholm, 1973, *Science*, 179, 792; "Radon Emanation from the Moon, Spatial and Temporal Variability" by Paul Gorenstein, Leon Golub & Paul Bjorkholm, 1974, *Moon*, 9, 129; "Recent outgassing from the lunar surface: the *Lunar Prospector* alpha particle spectrometer" by Stefanie L. Lawson, William C. Feldman, David J. Lawrence, Kurt R. Moore, Richard C. Elphic, Richard D. Belian & Sylvestre Maurice, 2005, *Journal of Geophysical Research*, 110, E9009. "In-orbit Performance of Alpha-Ray Detector (ARD) Onboard SELENE and the Early Results" by Katsuyuki Kinoshita, Yusuke Haruki, Masayuki Itoh, Takeshi Takashima, Takefumi Mitani, Kunishiro Mori, Jun Nishimura, Toshisuke Kashiwagi, Syouji Okuno & Kenji Yoshida, 2011, *Asia Oceania Geophysics Society*, Abstract PS10-A013. The Apollo Alpha Particle Spectrometer (APS) also flew on *Apollo 16*, but in a nearly equatorial orbit that covered only a small fraction of the lunar surface.

[27] The decay chain from uranium-238 to stable lead-206 is circuitous: $^{238}$U (half-life = $4.5\times10^9$ years) $\to_\alpha$ $^{245}$Th (25 days) $\to_\beta$ $^{234}$Pa (1.1 minute) $\to_\beta$ $^{234}$U ($2.3\times10^5$ years) $\to_\alpha$ $^{230}$Th (83000 years) $\to_\alpha$ $^{226}$Ra (1590 years) $\to_\alpha$ $^{222}$Rn (3.8 days) $\to_\alpha$ $^{218}$Po (3.1 minutes) $\to_\alpha$ $^{214}$Pb (27 minutes) $\to_\beta$ $^{214}$Bi (20 minutes) $\to_\beta$ $^{214}$Po (0.00015 second) $\to_\alpha$ $^{210}$Pb (22 years) $\to_\beta$ $^{210}$Bi (5 days) $\to_\beta$ $^{210}$Po (140 days) $\to_\alpha$ $^{206}$Pb, where U stands for uranium, Th for thorium, Pa for protactinium, Ra for radium, Rn for radon, Po for polonium, α for alpha emission (helium-4 nucleus), β for beta emission (an electron, but an anti-electron in some other cases). Of these only radon is gas at lunar temperatures. Note that betas alter element but not mass (not significantly); alphas alter both mass and element.

[28] "Low pressure radon diffusion: a laboratory study and its implications for lunar venting" by Larry Jay Friesen & John A.S. Adams, 1976, *Geochemica et Cosmochemica Acta*, 1976, 40, 375; "Radon Diffusion & Migration at Low Pressures, in the Laboratory and on the Moon" by Larry Jay Friesen, 1974, Ph.D. thesis, Rice University; "Model for Radon Diffusion through the Lunar Regolith" by L.J. Friesen & D. Heymann, 1976, *Moon*, 3, 461

[29] "Lunar Outgassing, Transient Phenomena & the Return to the Moon. II. Predictions & Tests for Outgassing/Regolith Interaction" by Arlin P.S. Crotts & Cameron Hummels, 2009, *Astrophysical Journal*, 707, 1506. We will discuss this model further below.

[30] "Chronological Catalog of Reported Lunar Events" by Barbara M. Middlehurst, Jaylee M. Burley, Patrick Moore & Barbara L. Welther, 1968 July, *NASA Technical Report TR R-277*, 64 pp.; "Lunar Transient Phenomena Catalog" by Winifred Sawtell Cameron, 1978 July, *NASA-TM-79399*, Greenbelt, Maryland: National Space Science Data Center/World Data Center A for Rockets & Satellites, Pub. 78-03, 116 pp.; "Lunar Transient Phenomena Catalog Extension" by Winifred Sawtell Cameron, 2006 July, unpublished, 152 pp. These contain 579, 1463 and 475 reports, respectively, with the Middlehurst et al. list nearly totally contained in Cameron 1978.

[31] "Detection of Radon Emission at the Edges of Lunar Maria with the Apollo Alpha-Particle Spectrometer" by Paul Gorenstein, Leon Golub & Paul Bjorkholm, 1974, *Science*, 183, 411.

[32] "Recent outgassing from the lunar surface: The *Lunar Prospector* Alpha Particle Spectrometer" (by Stephanie L. Lawson, William C. Feldman, David J. Lawrence, Kurt R. Moore, Richard C. Elphic, Richard D. Belian & Sylvestre Maurice, 2005, *Journal of Geophysical Research*, 110, E09009) shows the $^{210}$Po data, and ("Lunar Outgassing, Transient Phenomena & the Return to the Moon. I. Existing Data" by Arlin P.S. Crotts, 2008, *Astrophysical Journal*, 687, 692) the $^{210}$Po/mare edge correlation.

[33] "Lunar Outgassing, Transient Phenomena & the Return to the Moon. I. Existing Data" by Arlin P.S. Crotts, 2008, *Astrophysical Journal*, 687, 692.

[34] "Transient Lunar Phenomena, Deep Moonquakes & High-Frequency Teleseismic Events: Possible Connections" by Barbara M. Middlehurst, 1977, *Philosophical Transactions of Royal Society of London – A*, 285, 1327. "Shallow moonquakes - Argon release mechanism" by A.B. Binder, 1980, *Geophysical Research Letters*, 7, 1011; "Release of Radiogenic Gases from the Moon" by R.R. Hodges, Jr., 1977, *Physics of Earth & Planetary Interiors*, 14, 282.

[35] "Fluidization Phenomena and Possible Implications for the Origin of Lunar Crater" by A.A. Mills, 1969, *Nature*, 224, 863; and two articles by G.F.J. Garlick, G.A. Steigmann & W.E. Lamb, 1972, "Lunar dust – Effect of fluidization on reflective properties" *Nature*, 238, 13 & "Explanation of Transient Lunar Phenomena based on Lunar Samples Studies" *Nature*, 235, 39.

[36] "Lunar Outgassing, Transient Phenomena & the Return to the Moon. II. Predictions & Tests for Outgassing/Regolith Interactions" by Arlin P.S. Crotts & Cameron Hummels, 2009, *Astrophysical Journal*, 707, 1506.

[37] "Imaging of lunar surface maturity" by Paul G. Lucey, David T. Hawke, G. Jeffrey Taylor & B. Ray Hawke, 2000, *Journal of Geophysical Research*, 105, 20377.

[38] "The Optical Maturity of the Ejecta of Small Bright Rayed Lunar Craters" by J.A. Grier, A.S. McEwen, M. Milazzo, J.A. Hester & P.G. Lucey, 2000, *Lunar & Planetary Science Conference*, 31, 1950.

[39] "Lunar activity from recent gas release" by Peter H. Schultz, Matthew I. Staid & Carlé M. Peters, 2006, *Nature*, 444, 184; "High Resolution Imaging of Ina: Morphology, Relative Ages, Formation" by M.S. Robinson, P.C. Thomas, S.E. Braden, S.J. Lawrence, W.B. Garry & The LROC Team, 2010, *Lunar & Planetary Science Conference*, 41, 2592. See also "The geology and morphology of Ina" by P.L. Strain & F. El-Baz, 1980, *Lunar & Planetary Science Conference*, 11, 2437, and portions of the *Apollo 15 Preliminary Science Report*.

[40] "Search for Short-Term Changes in the Lunar Surface: Permanent Alterations Over Four Decades" by A.P.S. Crotts, 2011, *Lunar & Planetary Science Conference*, 42, 2600.

[41] For instance, outgassing events or meteoroid impacts might produce seismic signatures to be detected by lunar seismometers. While the four ALSEP seismometers were shut down on 1977 September 30, a smaller, future seismometer network might map the lunar interior if the place and time of seismic events can be localized. Imaging monitoring could supply this localization data.

[42] "Aristarchus Crater: Mapping of Impact Melt & Absolute Age Determination" by M. Zanetti, H. Hiesinger, C.H. van der Bogert, D. Reiss & B.L. Jolliff, 2011, *Lunar & Planetary Science Conference*, 42, 2330; "Ages & stratigraphy of mare basalts in Oceanus Procellarum, Mare Nubium, Mare Cognitum & Mare Insularum" by H. Hiesinger & J.W. Head III, 2003, *Journal of Geophysical Research*, 108, 5065 & citations therein.

[43] See "Compositional diversity and geologic insights of the Aristarchus crater from Moon Mineralogy Mapper data" by John F. Mustard, et al., 2011, *Journal of Geophysical Research*, 116, E00G12; "Clementine Observations of the Aristarchus Region of the Moon" by Alfred S. McEwen, Mark S. Robinson, Eric M. Eliason, Paul G. Lucey, Tom C. Duxbury & Paul D. Spudis, 1994, *Science*, 266, 1858.

[44] "Emergence of low relief terrain from shadow: an explanation for some TLP" by Raffaello Lena & Anthony Cook, 2004, *Journal of British Astronomical Association*, 114, 136. Winifred Cameron in the introduction to the 1978 TLP catalog states "Bartlett is an assiduous, experienced observer and has been rated high although most of his observations are of bluish phenomena on

Aristarchus, which may have their cause in terrestrial conditions, rather than lunar. If the user does not accept that explanation, then most of Bartlett's observations are very good."

[45] From now on I will use the notations "M68.N," "C78.N" and "C06.N" to refer to the Nth entry of the Middlehurst et al. 1968, Cameron 1978 and Cameron 2006 catalogs, respectively.

[46] From *The Historical Works of Gervase of Canterbury, Volume 1* by William Stubbs, 1879, London: Eyre & Spottiswoode, p. 276 (translated from Latin by Richmond Y. Hathorn, in Hartung 1976, below): "after sunset when the Moon had first become visible a marvelous phenomenon was witnessed by five or more men who were sitting facing the Moon. Now there was a bright new moon, and as usual in that phase its horns were tilted towards the east, and suddenly the upper horn split in two. From the midpoint of this division a flaming torch sprang up, spewing out, over a considerable distance fire, hot coals and sparks. Meanwhile the body of the Moon which was below writhed, as it were, in anxiety, and, to put it in the words of those who reported it to me and saw it with there own eyes, the Moon throbbed like a wounded snake. Afterwards it returned to its proper state. This phenomenon was repeated a dozen times or more, the flame assuming various twisting shapes at random and then returning to normal. Then after these transformations the Moon … along its whole length took on a blackish appearance. The present writer was given this report by men who saw it with their own eyes, and are prepared to stake their honor on an oath that they have made no additions or falsification in the above narrative." Gervase also gives plausible accounts of the conjunction of Mars and Jupiter on 1170 September 13 (*The Historical Works*, p. 221), a partial solar eclipse of 1178 September 13 (p. 21 & 279) and an apparent aurora on 1177 November 29 (p. 274).
Hartung hypothesized that Gervase reported an asteroid impact creating crater Giordano Bruno, but several papers criticize this idea. ("Was the Formation of a 20-km Diameter Impact Crater on the Moon Observed on June 18, 1178?" by Jack B. Hartung, 1976, *Meteoritics*, 11, 187) Crater counts superimposed on ejecta from Giordano Bruno indicate a crater age of 1 – 20 million years. ("Formation age of the lunar crater Giordano Bruno" by Tomokatsu Morota et al., 2010, *Meteoritic & Planetary Science*, 44, 1115) Also, the amount of material expected to hit Earth from such an event amounts to 10 million tonnes, and no evidence of such a meteor shower exists. ("Meteor storm evidence against the recent formation of lunar crater Giordano Bruno" by Paul Withers, 2001, *Meteoritic & Planetary Science*, 36, 525; also "On the Occurrence of Giordano Bruno Ejecta on the Earth" by Jack B. Hartung, 1981, *Lunar & Planetary Science Conference*, 12, 401; "Ejecta from lunar impacts: Where is it on earth?" by D.E. Gault & P.H. Schultz, 1991, *Meteoritics*, 26, 336) There is also some dispute about the interpretation of the report date. ("Author's Reply" by Paul Withers, 2001, *Meteoritic & Planetary Science*, 37, 466; "The 'lunar event' of AD 1178: A Canterbury Tale?" by Bradley E. Schaefer; Philip M. Bagnell, 1990, *Journal of British Astronomical Association*, 100, 211 – Note: Schaefer claims Gervase's solar eclipse date is seven days in error; nonetheless, it appears correct: NASA GSFC Eclipse Web Site, Solar Eclipse of 1178 September 13 – both report and post-diction on Julian September 13. He objects to identifying the AD 1177 event as an aurora since this was not solar maximum, but determining solar activity some 53 cycles before Cycle 0 of 1755 is problematic.)

[47] Porphyry is a dark red or purple rock. "Porphyritic" has more recently and independently come to mean igneous rocks with a texture of large crystals.

[48] *Selenotopographische Fragmente zur genauern Kenntniss der Mondfläche: ihrer erlittenen Veränderungen und Atmosphäre: sammt den dazu gehörigen Specialcharten und Zeichnungen* (*Selenotopographic fragments towards more exact knowledge of the lunar surface: its changes and sustained atmosphere together with accompanying special charts and drawings*) by Johann Hieronymus Schröter, 1791, Lilienthal: at author's expense, p. 533

[49] A detailed study is found in "Revisiting The 1963 Aristarchus Events" by B.E. O'Connell & A.C. Cook, 2012, *Journal of British Astronomical Association*, submitted.

[50] "The luminescence of the moon" by Zdeněk Kopal, 1965 May, *Scientific American*, 212, 28

[51] "A Suspected Partial Obscuration of the Floor of Alphonsus" by Dinsmore Alter, 1957, *Publications of Astronomical Society of Pacific*, 69, 158

[52] "A Photo-Visual Observation of an Impact of a Large Meteorite on the Moon" by Leon H. Stuart, 1956, *The Strolling Astronomer*, 10, 42; "Identification of the Lunar Flash of 1953 with a Fresh Crater on the Moon's Surface" by Bonnie J. Buratti & Lane L. Johnson, 2003, *Icarus*, 161, 192; "Lunar Flash Doesn't Pan Out" by J. Kelly Beatty, 2003 June, *Sky & Telescope*, 105, 24; "Optical Maturity Study of Stuart's Crater Candidate Impact" by D.T. Blewett & B.R. Hawke, 2004, *Lunar & Planetary Science Conference*, 35, 1098. See also "That Stuart Brilliant Flare & The Search for a New Lunar Crater" by Walter H. Haas, 2005, *The Strolling Astronomer*, 47, 46.

[53] "Langrenus: Transient Illuminations on the Moon" by Audouin Dollfus, 2000, *Icarus*, 146, 430; "First Results from Observations of the Moon by Means of a Polarimeter" by V.P. Dzhapiashvili & L.V. Ksanfomaliti, 1962, in *The Moon (IAU Symposium 14)*, eds. Z. Kopal & Z.K. Mikhailov, Waltham, Massachusetts: Academic, p. 463; "Some Results of Measurements of the Complete Stokes Vector for Details of the Lunar Surface (Nekotorye Rezul'taty Izmerenii Polnom Vektora Stoksa dlia Detalei Foverkhnosti Luny)" by Yu.N. Lipskii & M.M. Pospergelis, 1967, *Astronomicheskii Zhurnal*, 44, 410.

[54] "Observation of a Volcanic Process on the Moon" by Nikolai A. Kozyrev, 1959 February, *Sky & Telescope*, 18, 184; "The Kozyrev Observations of Alphonsus" by Dinsmore Alter, 1959, *Publications of Astronomical Society of Pacific*, 71, 46; "Volcanic Phenomena on the Moon" by Nikolai Kozyrev, 1963, *Nature*, 198, 979. The skepticism of Gerard Kuiper and his campaign to limit the effects of Kozyrev's lunar observations are detailed in "Evaluating Soviet Lunar Science in Cold War America" by Ronald E. Doel, 1992, *Osiris*, 7, 238. Kuiper saw Kozyrev's results as counter to the dominant American geologists' view of an inactive Moon, and was able to examine copies of his spectrographic plates sent by the editors of *Sky & Telescope*. (According to Doel, in a letter to the editor of *Sky & Telescope*, Kuiper implies that Kozyrev forged his data.) Kuiper's examination of the copies proved inconclusive, so he tried to obtain his own spectra at McDonald Observatory, also inconclusively. Kuiper then began an inquiry with Kozyrev's Soviet colleagues as his character. In the midst of this Urey, argued that there was no good reason to doubt Kozyrev's results, which further complicated the interdisciplinary politics of accepting Kozyrev's work. In late 1960 Kuiper attended a scientific meeting in Leningrad and was on that trip able to examine the original Kozyrev Alphonsus spectra for the first time. He changed his mind: he saw no reason to doubt Kozyrev's basic result, and upon returning to the U.S., he began
46

supporting the study of TLPs. (In fact Doel cites a member of Kuiper's staff as coining the original acronym for the occurrences: LTP, for "lunar transitory phenomena.")

[55] Brenner was in Lussinpiccolo, Croatia and Fauth in Lundstuhl, Germany, 750 km apart, so it is likely safe to assume they were independent. The reports were compared in a short article two years later: "Der Lichtschein im Plato" by Johann N. Krieger, 1897 March, *Sirius*, 30, 49.

[56] Communications transcript between *Apollo 11* and Mission Control, 1969 July 19; 8-digit code: days, hours, minutes, seconds after launch (Julian Date 2440419.0639 geocentric). Speakers are:
CC: Capsule Communicator (CAP COMM)  Bruce McCandless
CDR: Commander  Neil A. Armstrong
CMP: Command Module Pilot  Michael Collins
LMP: Lunar Module Pilot  Edwin E. Aldrin, Jr.

*03 04 57 07 CC:*     Roger. And we've got an observation you can make if you have some time up there. There's been some lunar transient events reported in the vicinity of Aristarchus. Over.
*03 04 57 28 LMP:* Roger. We just went into spacecraft darkness. Until then, why, we couldn't see a thing down below us. But now, with earthshine, the visibility is pretty fair. Looking back behind me, now, I can see the corona from where the Sun has just set. And we'll get out the map and see what we can find around Aristarchus.
*03 04 57 54 CDR:* We're coming upon Aristarchus right now –
*03 04 57 55 CC:*     – Okay. Aristarchus is at angle Echo 9 on your ATO chart. It's about 394 miles north of track. However, at your present altitude, which is about 167 nautical miles, it ought to be over - that is within view of your horizon: 23 degrees north, 47 west. Take a look and see if you see anything worth noting up there. Over.
*03 04 58 34 CDR:* Both looking.
*03 04 58 36 CC:*     Roger. Out.
*Note: here we skip 14 minutes and roughly 33 lines of dialog between the LMP, CDR and CC regarding navigation and communications channels. Aristarchus is mentioned only in terms of when Apollo 11 will be able to view it from orbit.*
*03 05 12 51 CMP:* Hey, Houston. I'm looking north up toward Aristarchus now, and I can't really tell at that distance whether I am really looking at Aristarchus, but there's an area that is considerably more illuminated than the surrounding area. It just has - seems to have a slight amount of fluorescence to it. A crater can be seen, and the area around the crater is quite bright.
*03 05 13 30 CC:*     Roger, 11. We copy.
*03 05 14*     (*NOTE: This is the time of the Pruss & Witte Aristarchus TLP report*)
*03 05 14 23 LMP:* Houston, Apollo 11. Looking up at the same area now and it does seem to be reflecting some of the earthshine. I'm not sure whether it was worked out to be about zero phase to – Well, at least there is one wall of the crater that seems to be more illuminated than the others, and that one – if we are lining up with the Earth correctly, does seem to put it about at zero phase. That area is definitely lighter than anything else that I could see out this window. I am not sure that I am really identifying any phosphorescence, but that definitely is lighter than anything else in the neighborhood.
*03 05 15 15 CC:*     11, this is Houston. Can you discern any difference in color of the illumination, and is that an inner or an outer wall from the crater? Over.
*03 05 15 34 CMP:* Roger. That's an inner wall of the crater.

*03 05 15 43 LMP:*  No, there doesn't appear to be any color involved in it, Bruce.
*03 05 15 47 CC:*     Roger. You said inner wall. Would that be the inner edge of the northern surface?
*03 05 16 00 CMP:*  I guess it would be the inner edge of the westnorthwest part, the part that would be more nearly normal if you were looking at it from the Earth.
*03 05 16 20 CC:*     11, Houston. Have you used the monocular on this? Over.
*03 05 16 28 LMP:*  Stand by one.
*03 05 17 59 LMP:*  Roger. Like you to know this quest for science has caused me to lose my E-memory program, it's in here somewhere, but I can't find it.
(*Note: the E-memory was the spacecrafts erasable memory that held temporarily programs for control of the spacecraft e.g., for guidance.*)
*03 05 18 08 CC:*     11, this is Houston. We're – we're hearing only a partial COMM. Say again please.
*03 05 18 20 CDR:*  I think ...
*03 05 18 41 CDR:*  Houston, we will give it a try if we have the opportunity on next - when we are not in the middle of lunch, and trying to find the monocular.
*03 05 18 51 CC:*     Roger. Copied you that time. Expect in the next REV you will probably be getting ready for LOI 2.
*03 05 19 09 CC:*     So, let's wind this up, and since we've got some other things to talk to you about in a few minutes. Over.

[57] At the time of these observations, the Moon's phase was 5.2 days past new, with Aristarchus in darkness 26° from the anti-solar point and 57° from the sub-Earth point on the Moon. The spacecraft was about 245 km in elevation above the lunar mean equatorial surface and some 750 km from the center of Aristarchus, which appeared inclined only 5° from edge-on. Only the north-northwestern part of the inner rim would be easily seen. At the time of the observation the phase angle was about 63°, whereas enhanced backscattering is significant only for angles of a few degrees. Backscatter does not explain these observations.

At mission elapsed time 03 05 14 (to the minute), Gail Pruss and Manfred Witte (at the Institute for Space Research, Bochum, West Germany) reported independently using a 150-mm telescope a 5 – 7 s brightening in Aristarchus (Cameron 1978 – we assign the longer timescale perhaps indicative of the *Apollo 11* report, otherwise the Pruss & White timescale is one of the shortest in the catalog). Note that *the Pruss and Witte report did not lead to the alert* by NASA, contrary to media reports e.g., "The Moon, A Giant Leap for Mankind," *Time*, 1969 July 25. It is unclear which event previous to this the Capsule Communicator indicates to *Apollo 11*. He probably refers to a pulsing glow in Aristarchus reported by Whelan from New Zealand 12.3 h earlier. That night has nine TLP reports in our sample, primarily from LION (see below), with seven involving Aristarchus over a 14 h interval, including independent reports (some with photographs) over 03 05 58 – 03 06 58 of Aristarchus being brighter than normal. There was no apparent attempt on Apollo 11 to observe Aristarchus on the next revolution, 2.15 h later in its initally wider and as yet uncircularized lunar orbit. In 2008 I asked the astronauts and neither recall a second attempt to observe this phenomenon. Collins has no clear recollection of the entire Aristarchus event, and Aldrin recalled no attempt to re-observe it. (The "LOI 2" burn – Lunar Orbit Insertion – at mission time 03 08 11 36 accomplished this circularization.)

The point of this extensive excerpt to is illustrate a few important issues at play in the TLP data set, particularly in the interval around the Apollo era. This is a unique example not only because of the setting, but because of the degree to which the information flow is documented. Is it a TLP report if observers are told to look at a specific area with special attention? Are the observers trained to distinguish the exceptional crater Aristarchus as a spatial anomaly rather than a temporal one in comparison to other craters?

Do perhaps observers sometimes dismiss real temporal anomalies because they have a mental model for normal appearance e.g., variations due to seeing – or in this case, the direct, 180° backscatter – that might be caused by less well-known effects? To what extent can simultaneous, independent reports differ in description and still constitute a confirmation? Is it significant that many earlier selenographers made careful, repeated observations with written records, or do more incidental observers provide useful reports as well? In the end this singular case is not a strong TLP report because even while simultaneous to another report for the same feature, it is unclear that the Apollo observers saw truly transient activity on relevant timescales.

[58] Based on the verbal descriptions, one can localize the Mattingly event to within about 10° on the lunar surface versus about 1° and 2° for the Schmitt and Evans events, respectively. The latter two reports occur during the relatively intense Geminid meteor shower (peaking December 14 versus times for the two reports of UT 1972 December 10 21:11 and 1972 Decemeber 11 22:28), and occur only 20° and 18° in longitude, respectively, from the point on the Moon's surface directly below the Geminid radiant. These two events occurred close to the point most likely to have been struck by Geminids. The Mattingly event (at UT 1972 April 21 19:01) occurs 142° in longitude from the leading point of the Moon's motion, in heliocentric coordinates, far from the most likely point for a meteoritic impact, but not conclusively so. (Note that the two *Apollo 17* events were 104° and 99° from the leading point, but could easily be Geminids.) The two *Apollo 17* reports during the Geminids might be explained by such impacts. The *Apollo 16* report has no such obvious explanation, but might also be due to impact.

[59] Several surveys have seen events e.g., five impacts during the Leonid meteor shower, three of them observed simultaneously, found by Ortiz et al. ("Optical detection of meteoroidal impacts on the Moon" by J.L. Ortiz, P.V. Sada, L.R. Bellot Rubio, F.J. Aceituno, J. Aceituno, P.J. Gutiérrez & U. Thiele, 2000, *Nature*, 405, 921; see also "Possibility of observing fall of meteorites on the moon from a station on the earth" by V.A. Anoshkin, G.G., Petrov & K.L. Mench, 1978, *Astromicheskii Vestnik*, 12, 216; "Making the photos of flashes on the Moon" by A.V. Arkhipov, 1991, *Zemlia i Vselennaya*, 3, 76; "Lunar Leonids: November 18th Lunar Impacts" by David Dunham, 1999, http://iota.jhuapl.edu/lunar_leonid/index.html.081299; "The first confirmed Perseid lunar impact flash" by Masahisa Yanagisawa, Kouji Ohnishi, Yuzaburo Takamura, Hiroshi Masuda, Yoshihito Sakai, Ida Miyoshi, Makoto Adachi & Masayuki Ishida, 2006, *Icarus*, 182, 489)

[60] "Rate and Distribution of Kilogram Lunar Impactors" by W.J. Cooke, R.M. Suggs, R.J. Suggs, W.R. Swift & N.P. Hollon, 2007, *Lunar & Planetary Science Conference*, 38, 1986; "The NASA Lunar Impact Monitoring Program" by Robert M. Suggs, William J. Cooke, Ronnie J. Suggs, Wesley R. Swift & Nicholas Hollon, 2008, *Earth, Moon & Planets*, 102, 293. A full sample is not yet published: "Flux of Kilogram-sized Meteoroids from Lunar Impact Monitoring" by Robert M. Suggs, W. Cooke, R. Suggs, H. McNamara, W. Swift, D. Moser & A. Diekmann, 2008, *Bulletin of American Astronomical Society DPS*, 40, 33.03; http://www.nasa.gov/centers/marshall/news/lunar/

ascribes to meteors; for though he does not suppose the moon to be surrounded with air, exactly like that which invests our globe, he thinks it probable that it may have an atmosphere of some kind, in which some of the elements of bodies, decompounded on its surface, may be suspended; and that some of the lunar mountains may emit nebulous vapours, not unlike the smoke of our volcanoes, which obscure and disguise the object seen through them. In the fourth book, we find a minute detail of the author's observations relative to those bright points, which have been seen on the moon's surface during eclipses, and, at other times, on her unenlightened part, and which some have supposed to be burning volcanoes. This opinion receives no countenance from M. Schroeter; who, after the most attentive examination of them, imagines that most of them must be ascribed to the light reflected from the earth to the dark part of the moon's disk, which returns it from the tops of its mountains, under various angles, and with different degrees of brightness. Some of these phenomena he suspects to be no more than optical illusions, arising from igneous meteors floating in our atmosphere, which happen to fall within the field of the telescope."

[69] "On the present State of the Question relative to Lunar Activity or Quiescence" by W.R. Birt, 1870, *Nature*, 2, 462: "From the time of Schröter, the question of change on the moon's surface has been more or less agitated. The *Selenotopographische Fragmente* contains numerous instances of what he considered to be changes of a temporary character, and a few of a more permanent nature, as the formation of new craters. It is, however, notorious that he failed to establish the fact of a decided change in any one instance; nor is this to be wondered at when we consider the paucity of the materials he had at his command. Notwithstanding the comparative neglect into which the observations recorded in the "Fragments" have fallen, and the judgments passed upon them by some of the best known selenographers, there can be no question that they embody the results of zealous and persevering attention to the moon's surface, and ought not to be passed over in the examination of any given spot, the history of which we are desirous of becoming acquainted with during the earliest period of descriptive observational selenography." Birt continues: "It is to Schmidt that we are indebted for one of the most important announcements bearing on the subject of lunar activity, that of a change in the crater Linné 'which,' says Mädler (Reports British Association, 1868, p. 517) 'has hitherto offered the only authentic example of an admitted change.'"

[70] For example: "Observations, Made by Several Astronomers, Domestic & Forrain, of the Late Eclipse of the Moon, on Septemb. 8, 1671, Here Delivered in the Languages, in Which they Were Communicated" by Street, et al., 1671, *Philosophical Transactions (1665-1678)*, 6, 2272; "Monsieur Auzout's Speculations of the Changes, Likely to be Discovered in the Earth & Moon, by Their Respective Inhabitants" by Auzout, 1665, *Philosophical Transactions (1665-1678)*, 1, 120.

[71] To concentrate on a particular feature, I choose Aristarchus and other relevant Aristarchus Plateau features that composes one-third of the raw report count (see Table 1 – Herodotus; Schroeter's or Schröter's Valley, or Vallis Schröteri; Cobra's Head or Cobra Head). To probe correlated observations statistically, I use an abstract/article search with a ADS Astronomy (and Physics) Query. Before 1930, in title or abstract, "volcano" AND (not OR) "aristarchus" produce no matches, whereas "aristarchus" alone produces 4 matches (not counting the ancient greek astronomer Aristarchus!) and "volcano" alone produces 13 articles relevant to the Moon. Note that this does not search the body of some longer articles, but perhaps satisfactorily reflects the attention that TLP-like claims might attract. Similar results are found for "gas," "atmosphere," "eruption," "flash," "cloud," "nebulosity," "mist," "geyser" and "vapo(u)r," with no matches for

any of these with "aristarchus." These results are summarized in Table A1.2. Likewise, replacing "aristarchus" with "herodotus" or various terms for Schröter's Valley or Cobra's Head produce no matches with the above terms for potentially TLP-like phenomena, or any articles describing TLPs. There is significant evidence that the majority, perhaps all, pre-1930 selenographers placed no special importance on possible, localized TLP activity, particularly in Aristarchus.

I also search the 1930-1955 literature for observer correlations in the literature (Table A1.3). The only citation involving ``change'' AND "aristarchus" is Haas (1938) referring to periodic changes appearance of the inner eastern wall of Aristarchus over nine-day intervals, hence not a TLP. The two other Aristarchus citations (Barcroft 1940, Barker 1942) concern the same subject and were evidently written in reaction to Haas (1938). This is the type of statistical correlation between events we would guard against if they involved TLPs. ("Lunar Changes in the Crater Aristarchus" by Walter H. Haas, 1938, *Popular Astronomy*, 46, 135; "The bands of Aristarchus" by David P. Barcroft, 1940, *Popular Astronomy*, 48, 302; "The Bands of Aristarchus" by Robert Barker, 1942, *Popular Astronomy*, 50, 192)

Over 1930–1955, the 26 citations involving "moon" AND "change" include six dealing with lunar appearance changes, all over days or weeks, over many lunar features, not concentrating on any strong TLP sites (except Haas 1938). These include Atlas, Billy, Crüger, Endymion, Eratosthenes, Eudoxus, Furnerius, Grimaldi, Hercules, Linné, Macrobius, Mare Crisium, Messier, Phocylides, Pickering, Pico/Pico B, Plato, Riccioli, Rocca, Snellius, Stevinus, and Theophilus. The only citation involving "moon" AND "transient" is irrelevant.

The year 1956 delineates the period after which more TLPs are reported, after Alter (1957) [51] and Kozyrev (1959) [54], inspiring further observations in a cascade through the catalog. This wreaks havoc with our ability to evaluate TLP observational biases, and implies that later reports cannot be considered isolated entities. Post-1955 citation correlations show this (Table A1.4). In 1956–1968, all 10 Aristarchus references, seven involving TLPs, lead back to Kozyrev's reports. Of these seven, five involve TLP-related search terms that we have used. During this period TLPs become fixed in people's minds in connection with particular features such as Aristarchus.

[72] "Lunar Changes in the Crater Aristarchus" by Walter H. Haas, *Popular Astronomy*, 46, 135.

[73] "The Lunar Craters Aristarchus & Herodotus" by W.R. Birt, 1870, *Astronomical register*, 8, 271; "The Lunar Craters Aristarchus & Herodotus" by H. Michell Whitley, 1870, *Astronomical register*, 8, 193.

[74] *Are The Planets Inhabited?* by E.W. Maunder, 1913, London: Harper.

[75] "Are there at present Active Volcanos upon the Moon?'" by William H. Pickering, 1892, *The Observatory*, 15, 250. His book: *The Moon: A Summary of the Existing Knowledge of Our Satellite, With a Complete Photographic Atlas* by William Henry Pickering, 1903, New York: Doubleday.

[76] "Selenographic Notes, February, 1884: Aristarchus & Herodotus" by Thos. Gwyn Elger, *Astronomical Register*, 22, 39.

[77] "The Moon in Ultra-Violet Light: Spectro-Selenography" by R.W. Wood, 1910, *Monthly Notices of Royal Astronomical Society*, 70, 226; reference to "Wood's Spot" – "Lunar Color Boundaries and Their Relationship to Topographic Features: A Preliminary Survey" Ewen A. Whitaker, 1972, in *Conference on Lunar Geophysics*, (Houston: Lunar Science Institute), p. 348.

[78] Since each pixel can be labeled with the name of the feature(s) identified by the observers in the reports that filled that pixel, we can list the corrected count for each feature or group of features. Within each pixel, we re-evaluate particular features to see if TLPs from the three samples truly correspond geographically. If TLPs occur in the same named feature (and we include any positional information available), or within a 50 km radius of each other, or within 1.5 times the radius of the named crater, whichever is larger, we retain this as a match. The latter is a rejection consideration in less than 10% of the cases.

**Figure 1:**

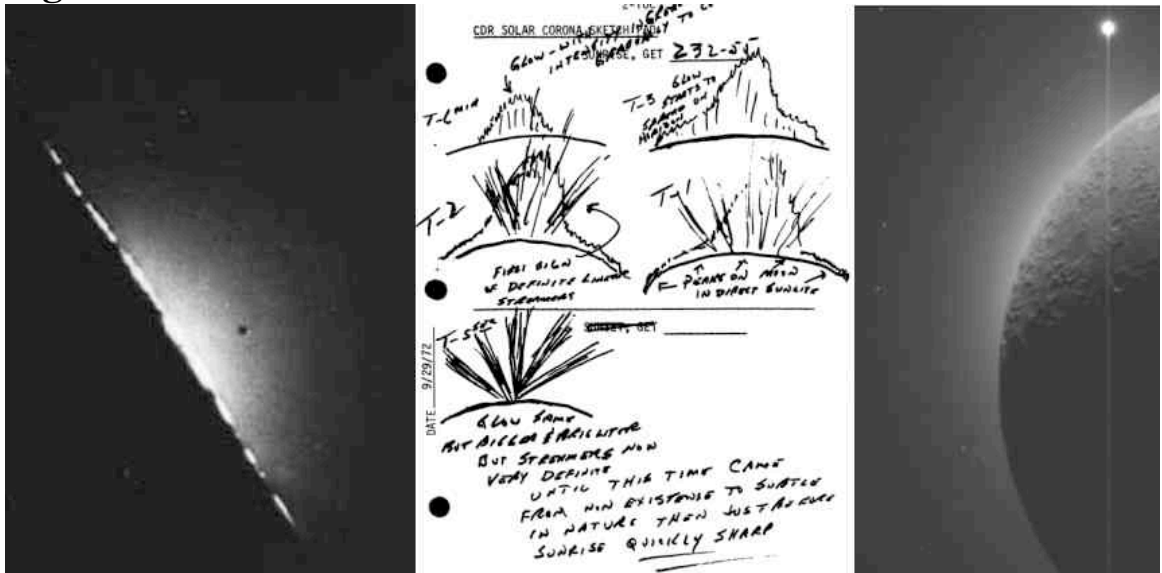

Views of dust-scattered light over Moon's surface. **Left**: *Surveyor 6* took this TV image of the lunar horizon on 1967 November 24 about 40 minutes after sunset. Note the line of bright patches on the horizon, now interpreted as scattering by dust clouds levitated by photo-electric charging of the soil by solar ultraviolet light; **Center**: sketches made by Gene Cernan on *Apollo 17* as the sun rose over the Moon as seen from orbit on 1972 December 16 around 22:28 UTC, at 6, 5, 2, 1 and $^1/_{12}$ minutes before sunrise. Note the extension of light along the horizon a few minutes before sunrise, followed by bright streamers immediately before sunrise. The rapidly changing structure, even in a matter of seconds, indicated an effect of local lunar dust; **Right**: the Moon occulting the Sun (with Venus near the top of the image creating a vertical signal saturation streak), taken by the *Clementine* star tracker camera on 1994 April 18 at 01:12:52 UTC, about 4500 kilometers from the Moon. One can see diffuse light from dust scattering near the Moon's surface, but far above it is mixed with zodiacal light from dust in solar orbit and seems to be weaker than the glow seen by Cernan. (Photos 67-H-1642, S-83-15138, LBA5881Z, courtesy of NASA)

**Figure 2:**

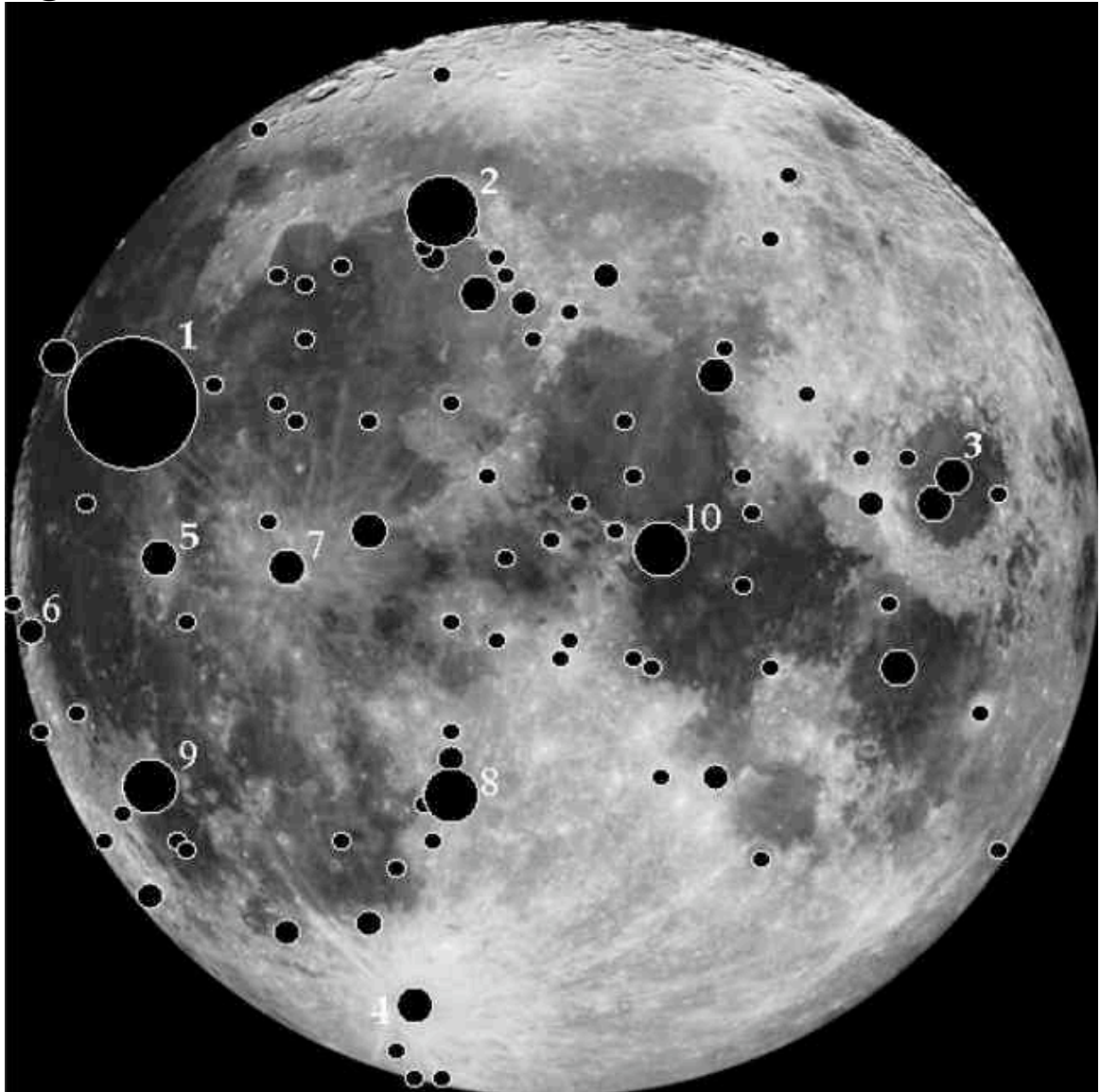

Distribution of TLP report loci as catalogued in Middlehurst et al. (1968), with the exception of a minority of cases that are rejected for the reasons detailed in the text. The size of the symbols encodes the number of reports per features, as listed in Table 1. Marked features include: 1) Aristarchus (including Schröter's Valley, Cobra's Head and Herotus), 2) Plato, 3) Mare Crisium, 4) Tycho, 5) Kepler, 6) Grimaldi, 7) Copernicus, 8) Alphonsus, 9) Gassendi, and 10) Ross D. (Photo *Galileo* PIA00405 courtesy of NASA)

**Figure 3:**

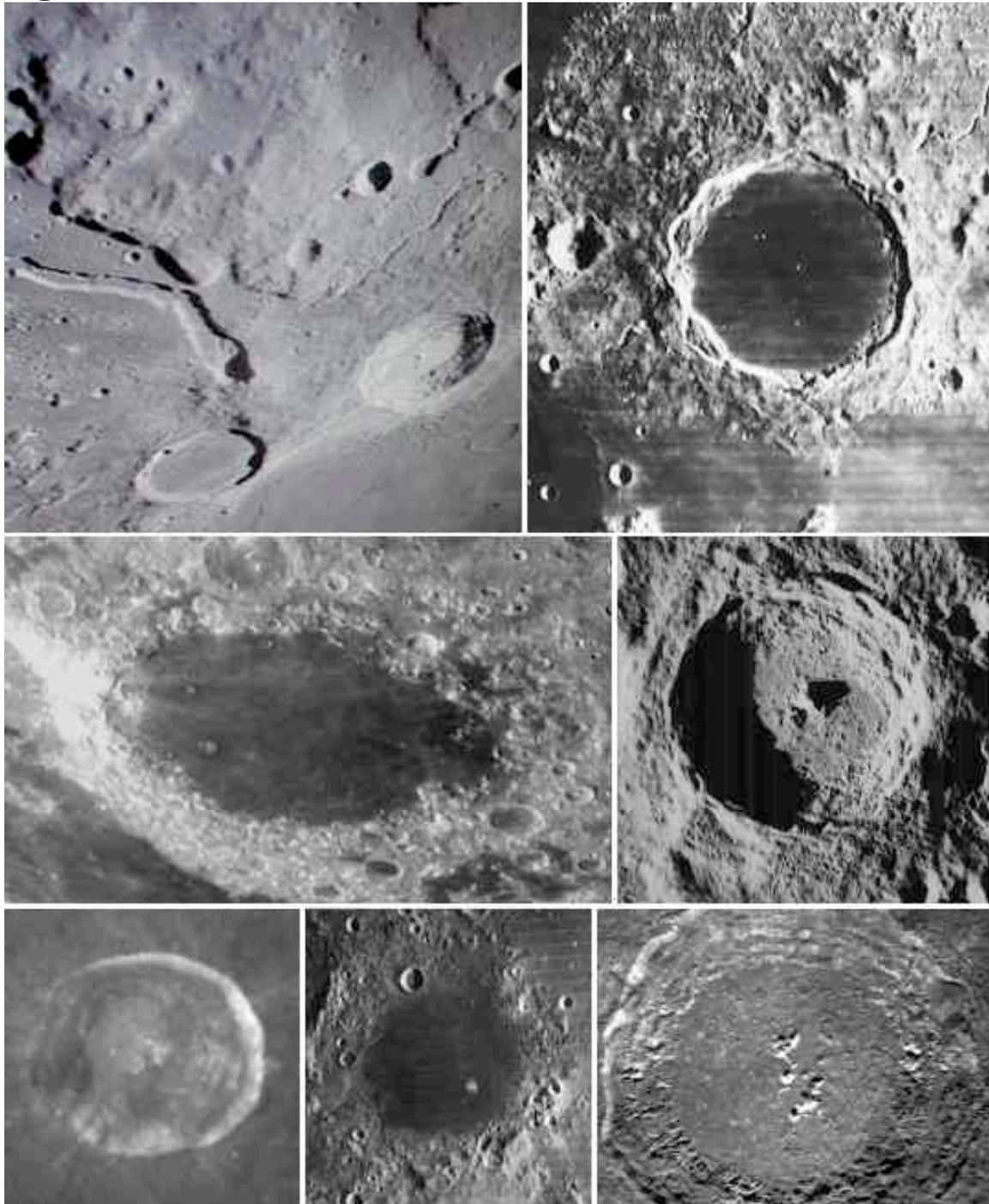

Various lunar missions' views of robust TLP report activity sites: 1) southeast Aristarchus Plateau: Herodotus in lower (distant) left, Aristarchus crater in lower left, Cobra Head above Herodotus extending into Schröter's Valley at left. Notice many other rilles running off the plateau. Image is foreshortened, roughly 160 kilometers on a side; 2)

Plato.  Note the rilles at right.  Image is 230 kilometers wide; 3) Mare Crisium. Foreshortened image is 760 kilometers across; 4) Tycho (image 130 kilometers wide); 5) Kepler (42 kilometers); 6) Grimaldi (280 kilometers); 7) Copernicus (95 kilometers).  Aristarchus, Copernicus, Kepler & Tycho are similar in appearance; this figure shows the effects of illumination angle. (Photos courtesy of NASA)

**Figure 4:**

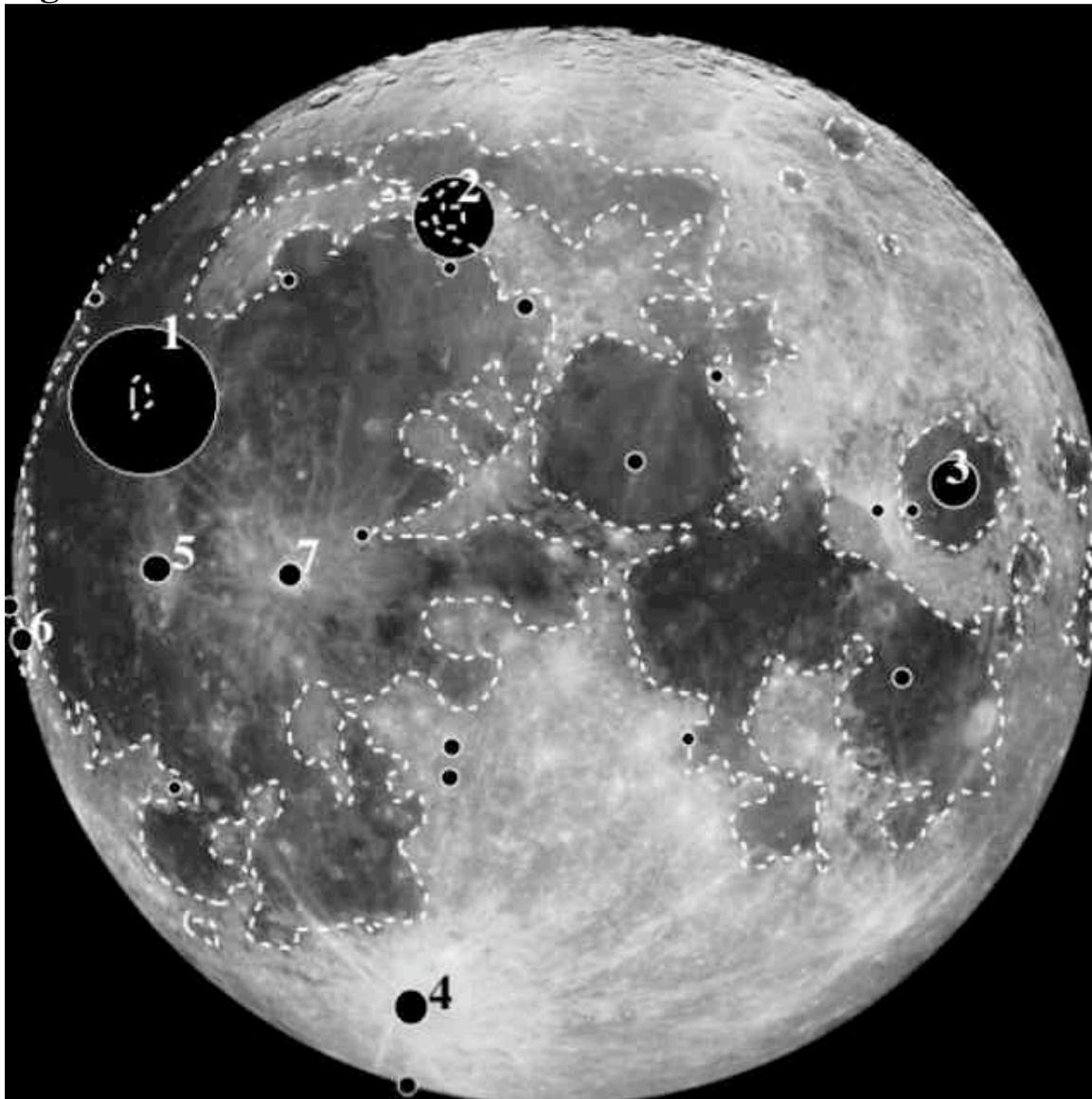

Distribution of robust TLP sites, as explained in the text and found in Table A.9. The size of the symbols encodes the fraction per features of the total reports. Symbols correspond to 1) Aristarchus (including Schröter's Valley, Cobra's Head and Herotus), 2) Plato. 3) Mare Crisium, 4) Tycho, 5) Kepler, 6) Grimaldi, 7) Copernicus. The dashed curved contour is the adopted boundary between mare and highlands, as explained in the text. (Photo *Galileo* PIA00405 courtesy of NASA)

**Figure 5:**

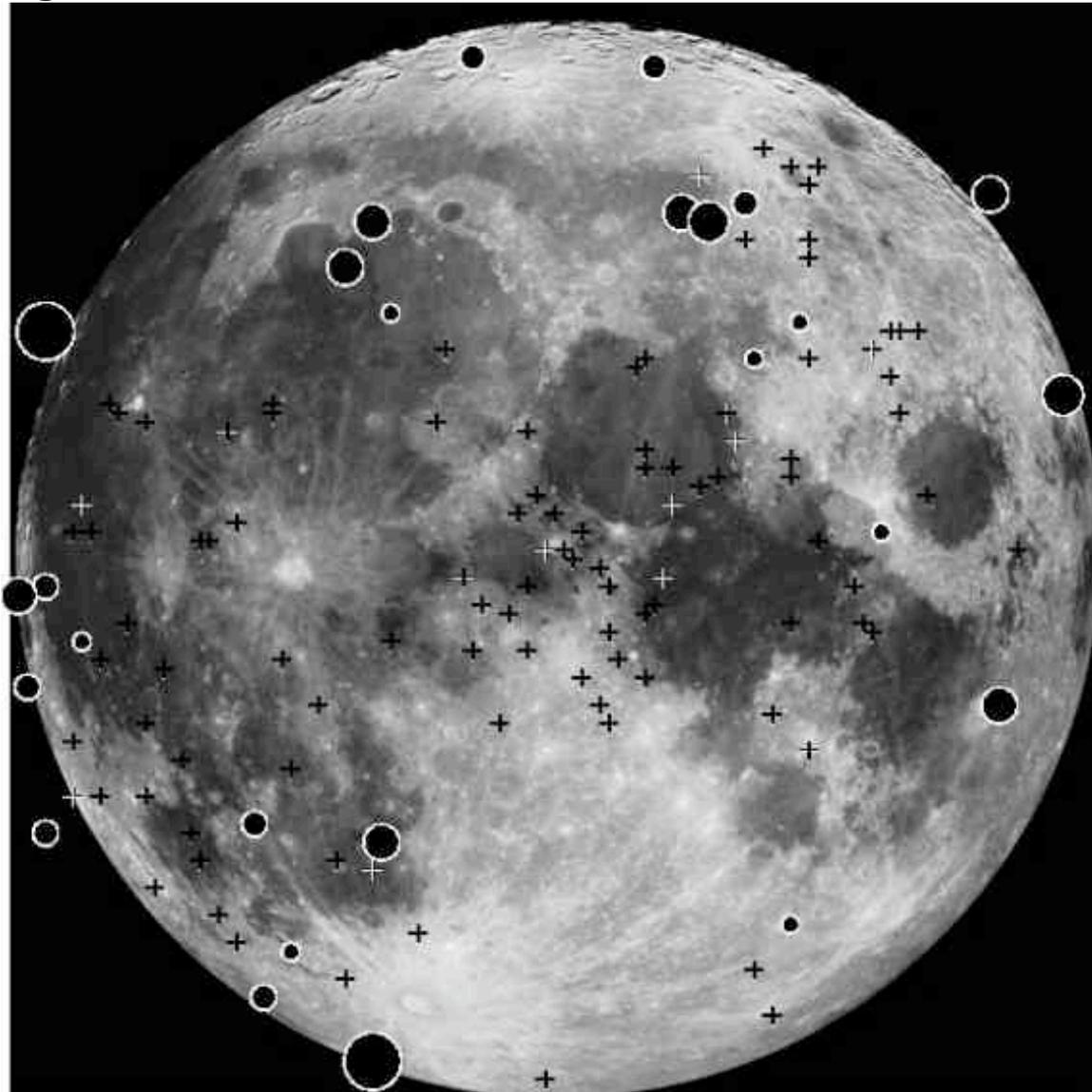

The distribution of moonquakes, primarily on the Near Side (where detections concentrate due to the nearside placement of seismometers). Shallow moonquakes (less than 700 km deep) are marked by filled circles, some off the lunar surface by an amount equal to the distance they wrap onto the Far Side. The diameter of shallow quake markers scale with their magnitudes. Deep moonquakes are shown as small crosses. Their correlation with mare edges is obvious. While the chance of shallow moonquakes' correlation with mare edges is random has a probability of 0.0002, the deep moonquakes almost certainly correlate with mare edges, to a certainty of one part in $10^{14}$.

**Figure 6:**

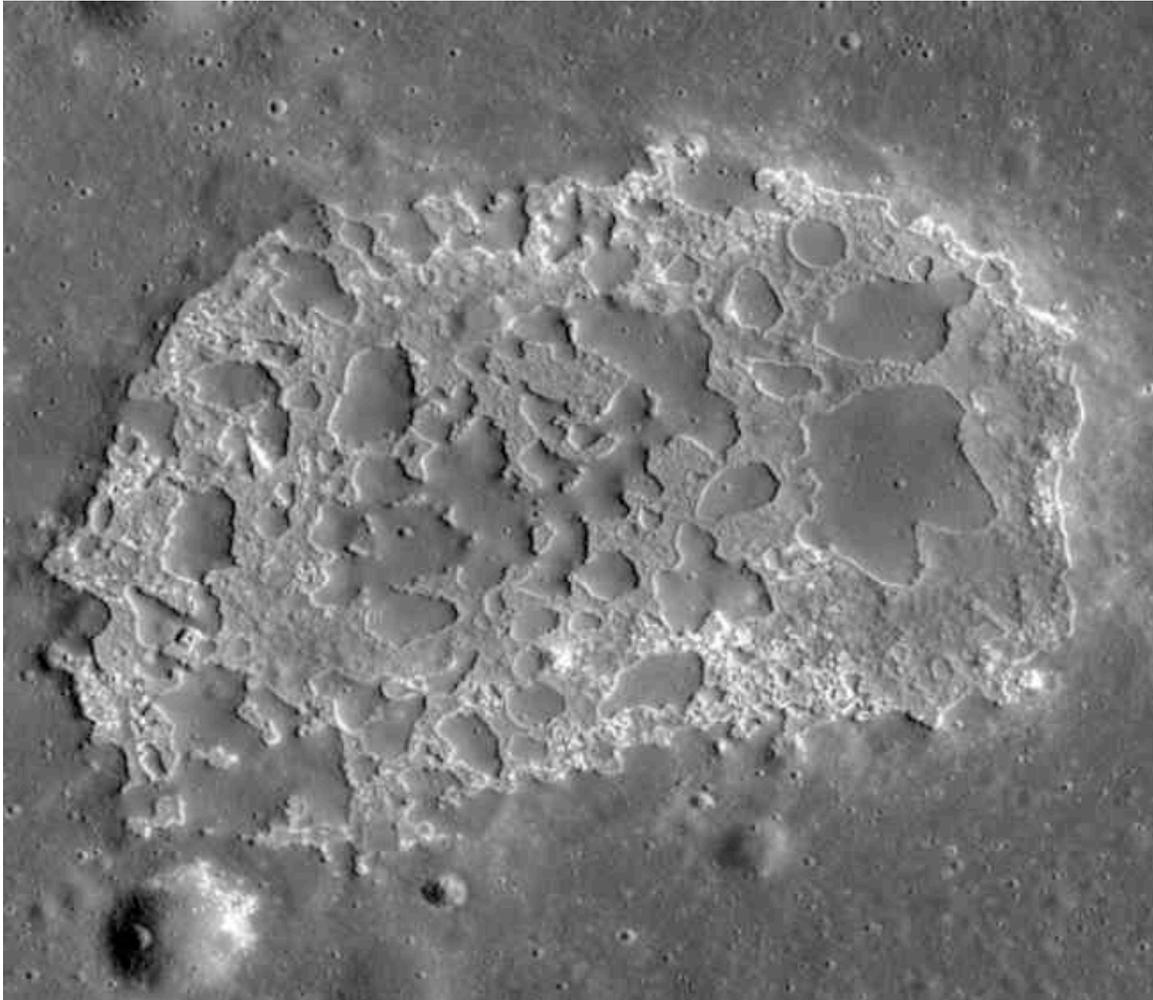

The Ina feature is a depression not excavated by impact. It is 3 km wide and 60 meters deep, and has an age measured by OMAT of a few million years or less. The lighter areas are deeper than the grey areas they surround, and some of the grey areas rise about 40 meters, approaching the level of the surrounding regolith surface. (NASA/ASU/GSFC LROC-NAC image M137929856R)

**Figure 7:**

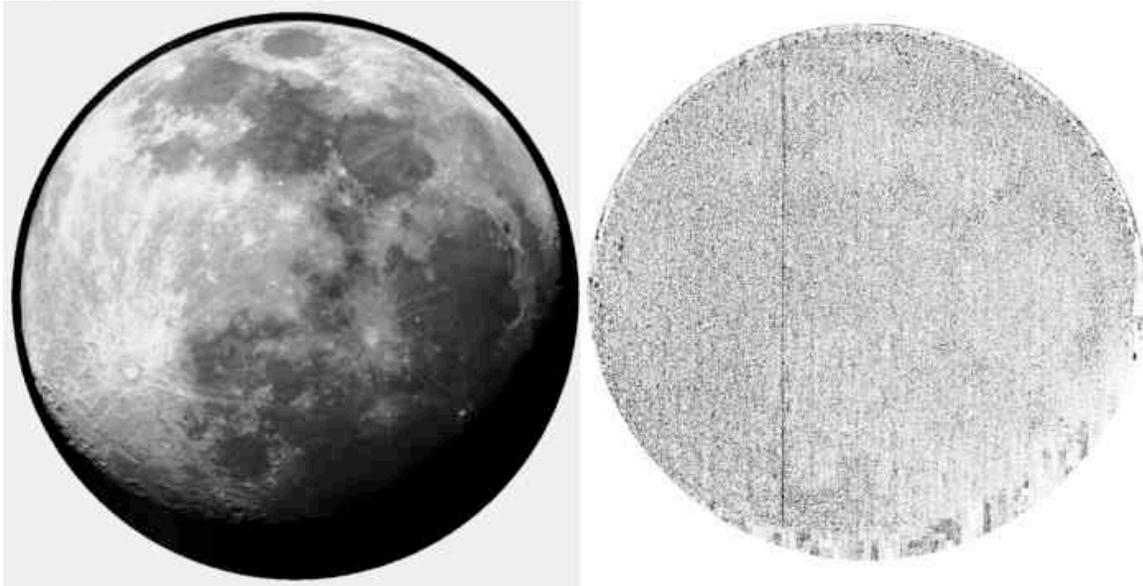

**Left:** Flat-field, dark-corrected but otherwise "raw" image of a typical lunar Near Side image obtained by our robotic imaging monitor. (The image is trimmed to a standard circular region). **Right:** The difference in signal between the image above and similar one obtained five minutes later. The noise in the residual signal is essentially at the photon shot-noise limit. Because of a slight error in the photometric calibration between the two images, there is a very slight ghost of high-contrast global features, especially Imbrium, Humorum and the eastern maria. Note that even bright smaller features e.g., Tycho left and below center, are subtracted nearly identically. Even subtle features not apparent in the image above e.g., an image column acting slightly non-linearly, just left of center, becomes readily apparent. There are also errors along the lunar limb due to the rapid gradient in signal level.

**Figure 8:**

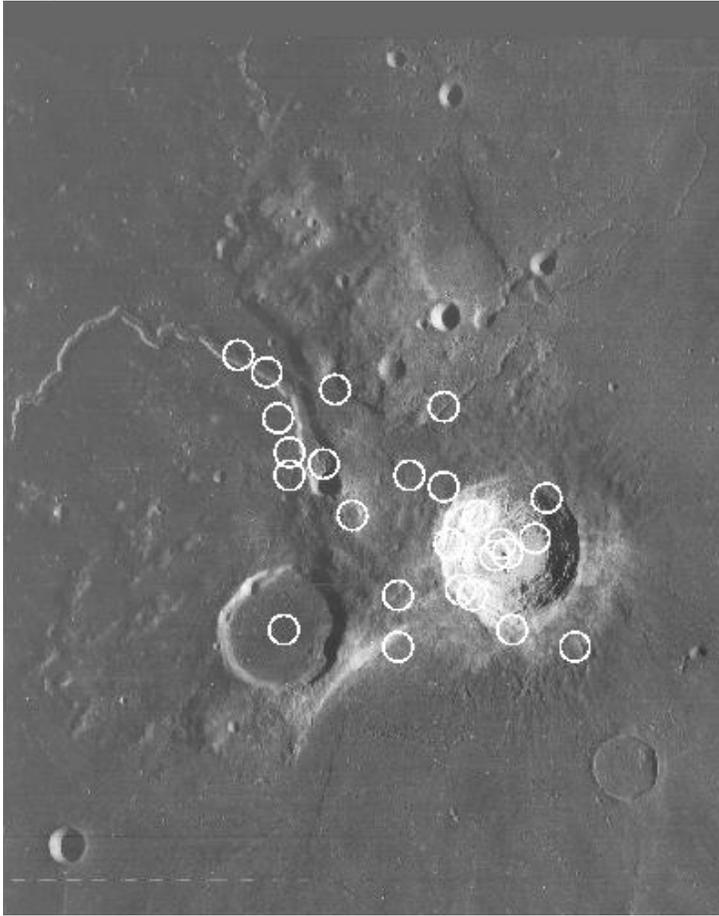

A *Lunar Orbiter* image of Aristarchus (bright crater, right of center), Herodotus (below right of center), Vallis Schröteri (extending from center to left edge), and the southeastern quadrant of the Aristarchus Plateau (the southern edge running horizontally below Aristarchus and eastern edge vertically to the right). The white circle indicate TLP reports from the historical record before 1957 that can be localized to within 10 km. (The view is approximately 220 km across.) Note the concentration of reports within the crater Aristarchus and on the Plateau adjacent and running down Vallis Schröteri.

**Figure 9:**

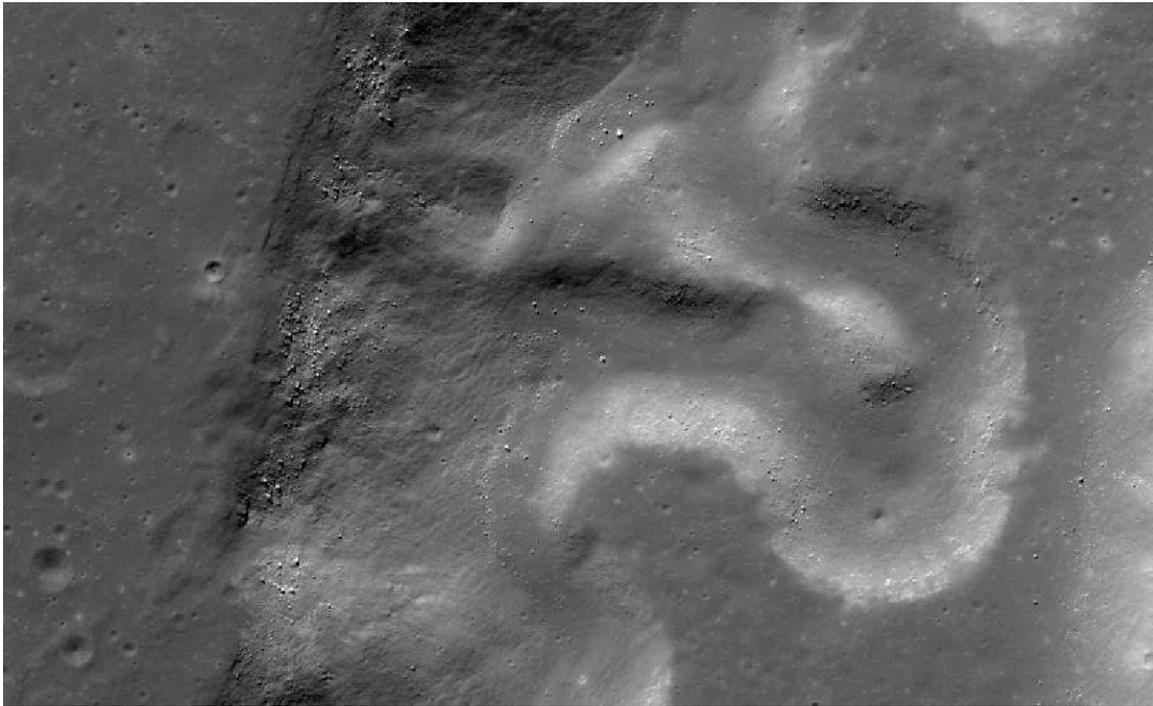

View of Vallis Schröteri seen from LROC-NAC (image M104848322L), showing, from the left, the surface typical of the Aristarchus Plateau, with high density of cratering and dark mantling. About one-quarter of the way from the left one encounters crest of the outer rille showing mass wasting and lower crater density, sloping down to the level of the outer rille just left of center. The left half of the outer rille is actually occupied by the inner rille, which its oxbow, meandering shape. The floor of the outer rille then continues to the right edge, where it meets the base of upslope back to the Plateau at bottom right. Mote that the floor on the inner rille, at the bottom of the oxbows, are lightly cratered, while the cratering density on the floor of the outer rille in intermediate between the inner rille and Plateau. The field of view is 4.8 km across.

**Figure A1.1**

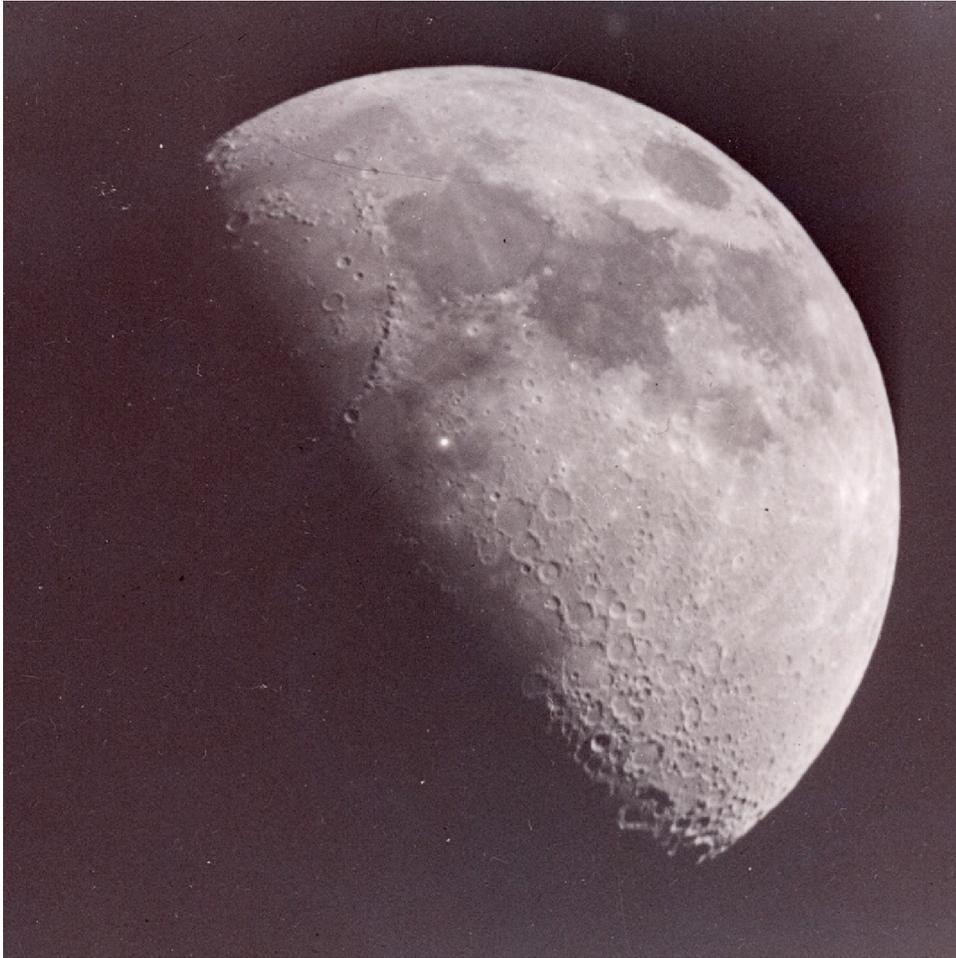

Leon Stuart recorded this TLP on 1953 November 15 in which he saw a bright spot on the Moon (near center, in Sinus Medii), recorded this photograph, but then noticed that the spot had disappeared before taking a second photograph. He estimates that the event lasted about 8 s. (Photo used with permission of Jerry Stuart)

**Figure A1.2:**

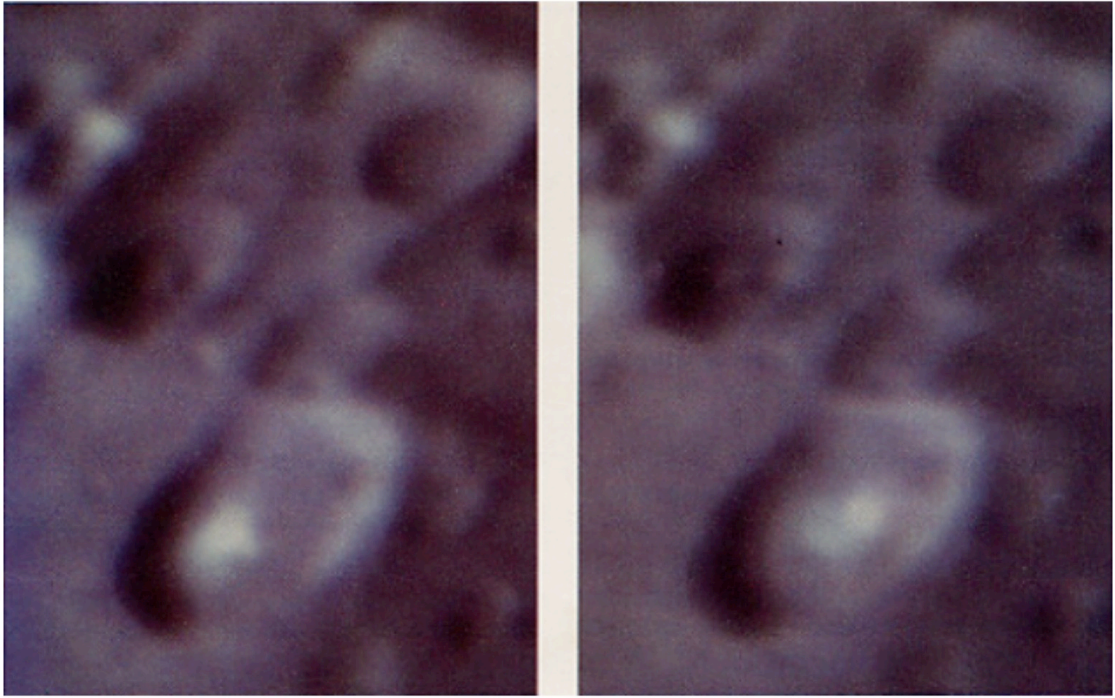

G. Slayton recorded these two images, about 20 minutes apart on 1981 September 5, showing a luminous spot moving (at about 30 km/hour) across the floor of Piticus (Photograph used with permission of W. Cameron)

**Figure A1.3**

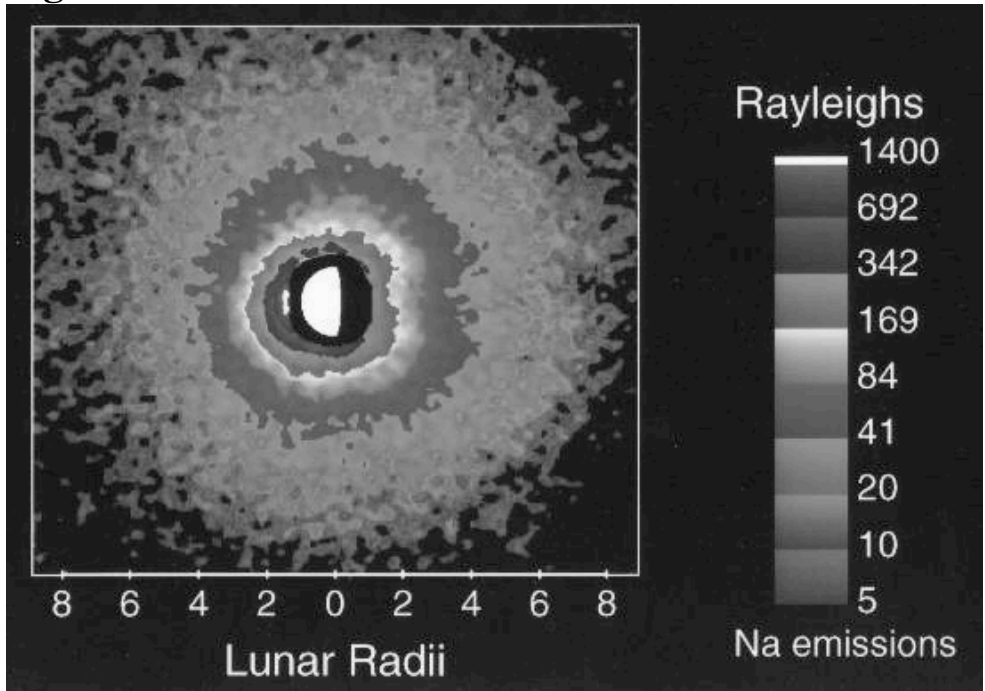

Moon's extended sodium atmosphere as observed at McDonald Observatory 1991 September 30 (Courtesy of Dr. Michael Mendillo & Boston University)